\documentclass{article}

\usepackage[main, final]{neurips_2026}

\usepackage[utf8]{inputenc} % allow utf-8 input
\usepackage[T1]{fontenc}    % use 8-bit T1 fonts
\usepackage{hyperref}       % hyperlinks
\usepackage{url}            % simple URL typesetting
\usepackage{booktabs}       % professional-quality tables
\usepackage{amsfonts}       % blackboard math symbols
\usepackage{nicefrac}       % compact symbols for 1/2, etc.
\usepackage{microtype}      % microtypography
\usepackage{xcolor}         % colors

\usepackage[english]{babel}
\usepackage{nicematrix}

\usepackage{amsmath, amssymb, amsthm, mathtools, mathrsfs}
\usepackage{dsfont}             % \mathds{R} 等
\usepackage{graphicx}
\usepackage[export]{adjustbox}
\usepackage{caption, subcaption}
\usepackage{wrapfig}
\usepackage{booktabs}
\usepackage{array, multirow}
\usepackage{tabularx, xltabular}
\usepackage{makecell}
\usepackage{nicematrix}

\usepackage{tikz}
\usepackage{pgfplots}
\pgfplotsset{compat=1.18}   % 建议显式设置版本

\usepackage{algorithm}
\usepackage{algpseudocodex}

\usepackage{microtype}
\usepackage{enumitem}
\usepackage{setspace}
\usepackage{indentfirst}
\usepackage{numprint}
\usepackage{balance}
\usepackage[normalem]{ulem}
\usepackage{soul}
\usepackage{romannum}
\usepackage{xspace}

\usepackage{xcolor}
\usepackage{tcolorbox}
\tcbuselibrary{skins, breakable, theorems}

\usepackage{afterpage}
\usepackage{textcomp}
\usepackage{verbatim}

\usepackage{pifont}
\newcommand{\cmark}{\ding{51}}  % 绿色对勾
\newcommand{\xmark}{\ding{55}}  % 红色叉号

\definecolor{promptDarkGreen}{RGB}{0, 128, 0}
\definecolor{promptLightGreen}{RGB}{240, 250, 240}
\definecolor{inputGray}{gray}{0.30}

\newtcolorbox{examplebox}[1][]{
    enhanced, colback=gray!10, colframe=gray!75,
    sharp corners, boxrule=0.5pt, title=#1
}

\newtcolorbox{promptbox}[1]{
    enhanced, breakable,
    colback=promptLightGreen, colframe=promptDarkGreen,
    coltitle=white, fonttitle=\bfseries\large,
    title={#1},
    sharp corners=south,
    boxrule=1mm, arc=1mm,
    width=\linewidth
}

\newtcolorbox{userinputbox}{
    colback=white, colframe=gray!30,
    boxrule=0.5mm, sharp corners,
    left=2mm, right=2mm, top=2mm, bottom=2mm
}
\newcommand{\name}{$\mathtt{AgentGEO}$\xspace}
\newcommand{\benchmark}{$\mathbf{MIMIQ}$\xspace}

\title{Diagnosing and Repairing Citation Failures in Generative Engine Optimization}

\author{
  Zhihua Tian$^{1}$\thanks{Equal contribution.}\hspace{0.5em}\thanks{Correspondence to: zhihuat@vt.edu.}
  \quad
  Yuhan Chen$^{1}$\footnotemark[1]
  \quad
  Yao Tang$^{2}$\footnotemark[1]
  \quad
  Jian Liu$^{2}$
  \quad
  Ruoxi Jia$^{1}$ \\
  \\
  $^{1}$Virginia Tech \qquad
  $^{2}$Zhejiang University
}

\begin{document}

\maketitle

\begin{abstract}
  Generative Engine Optimization (GEO) aims to improve content visibility in AI-generated responses. However, existing methods measure contribution—how much a document influences a response—rather than citation, the mechanism that actually drives traffic back to creators. Also, these methods apply generic rewriting rules uniformly, failing to diagnose why individual documents are not cited.
This paper introduces a diagnostic approach to GEO that asks why a document fails to be cited and intervenes accordingly. We develop a unified framework comprising: (1) the first taxonomy of citation failure modes spanning different stages of a citation pipeline; (2) \name, an agentic system that diagnoses failures using this taxonomy, selects targeted repairs from a corresponding tool library, and iterates until citation is achieved; and (3) a document-centric benchmark evaluating whether optimizations generalize across held-out queries.
\name achieves 15-40\% relative improvement in citation rates across engines and citation methods, while modifying only 5\% of content, compared to 25\% for baselines. Our analysis reveals that generic optimization can harm long-tail content and some documents face challenges that optimization alone cannot fully address—findings with implications for equitable visibility in AI-mediated information access.
Project page: \url{https://zhihuat.github.io/agentgeo/}.

\end{abstract}

\section{Introduction}
The rise of generative engines has sparked significant concern among content creators about traffic loss~\cite{economist2025ai}. Unlike traditional search engines that drive users to click through to websites, generative engines synthesize answers directly, potentially reducing the organic traffic that sustains the creator economy. In response, major platforms have introduced \emph{citation} mechanisms—Google AI Overview displays source links alongside summaries~\cite{google2025aioverviews}, Perplexity provides inline citations~\cite{PerplexityAI2026}, and ChatGPT's search mode includes expandable references~\cite{OpenAI2024ChatGPTSearch}. These citations are now the primary pathway from synthesized answers back to original webpages.

This pathway, however, is narrow. A recent analysis of 68,879 Google searches reports that only 1\% of users who encounter AI summaries click cited sources, compared to 15\% who click traditional results when no AI summary appears~\cite{chapekis2025google}. Yet the traffic that does arrive could convert at substantially higher rates—one analysis found generative search visitors convert 23 times better than traditional organic visitors~\cite{stox2025ahrefs}. For content creators, citation thus becomes a critical gateway: although click-through is rare, the visitors who do arrive are disproportionately valuable, making the difference between being cited and being overlooked all the more consequential.

\textbf{From Contribution to Citation.} These dynamics have motivated growing interest in Generative Engine Optimization (GEO): modifying web content to improve visibility in AI-generated responses. Existing work evaluates visibility with \emph{contribution} metrics such as Word Count, which measure the fraction of a response attributed to a source~\cite{aggarwal2024geo}. While useful, these continuous metrics conflate two factors: whether a webpage is cited at all and, conditional on citation, how much of the response is attributed to it. For content creators, the critical barrier is non-citation: no citation means no referral opportunity. We therefore focus on a complementary, citation-first objective that isolates the binary event of being cited. In our experiments, 43\% of topically relevant webpages receive no citation under baseline conditions. For these webpages, the question is not ``how much am I cited?'' but ``why am I not cited at all?'' Understanding and repairing these failures is an underexplored problem with direct implications for the sustainability of web content creation.

\begin{wrapfigure}[19]{r}{0.5\linewidth}
    \centering
    \includegraphics[width=\linewidth]{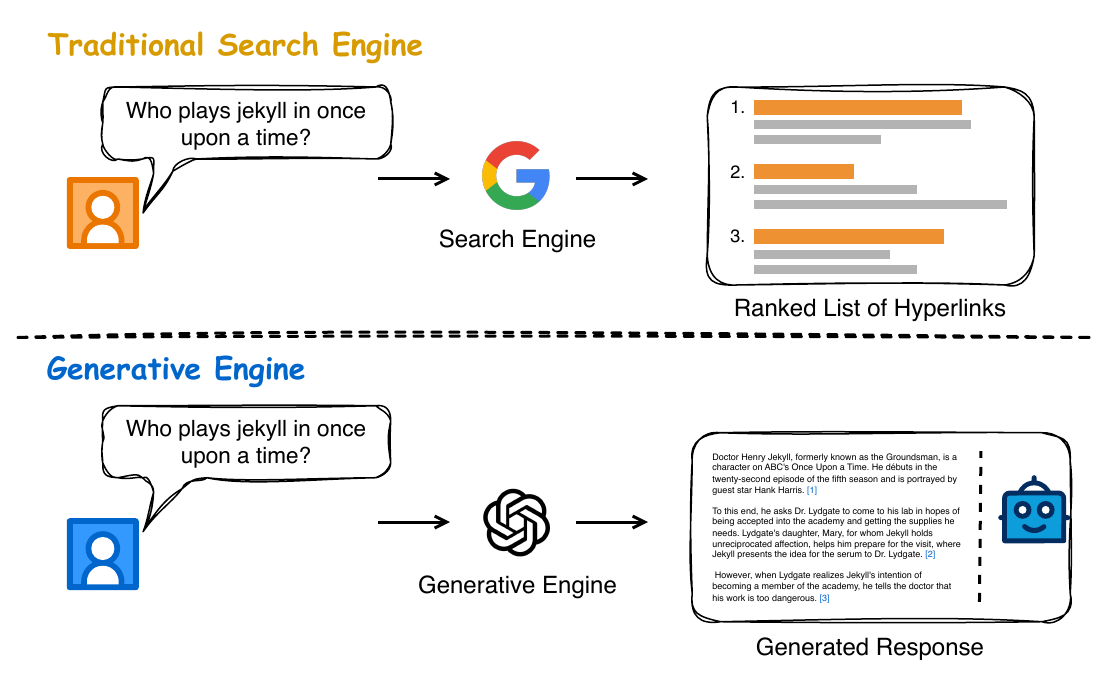}
    \caption{Comparison between traditional search engines and LLM-powered generative engines. Traditional search engines return a ranked list of hyperlinks, while generative engines synthesize answers from retrieved content and provide citations to source webpages.} \label{fig:search_engine_comparision}
    \label{fig:wrap}
\end{wrapfigure}

\textbf{Limitations of Existing Methods.} Current GEO methods apply generic preference rules to text uniformly: add statistics, adopt authoritative tone, improve fluency~\cite{aggarwal2024geo,wu2025generative}. This approach fails because citation failures are heterogeneous, spanning multiple pipeline stages. A webpage may fail at \emph{fetching} (malformed HTML, JavaScript dependencies), at \emph{parsing} (content truncated or buried below boilerplate), or at \emph{generation} (missing entities, inferior information density versus competitors). Each failure mode requires a different intervention, yet generic rules cannot diagnose which stage failed, nor can they address upstream failures that occur before text is even processed.

\textbf{Contribution 1: A Diagnostic Approach to GEO.} Motivated by these observations, we reframe GEO from ``what generic improvements might help?'' to ``why did this webpage fail to be cited for the relevant queries?'' We instantiate this perspective in AgentGEO, an agentic framework for iterative diagnosis and repair. Given a webpage that fails to be cited, AgentGEO compares it against cited competitors to identify the most likely disadvantage, selects an appropriate tool from a library of specialized interventions spanning the entire citation pipeline, and verifies whether citation is achieved. If not, it re-diagnoses and iterates, with a memory mechanism to track prior attempts.

\begin{figure*}[ht!]
    \centering
    \includegraphics[width=0.9\linewidth]{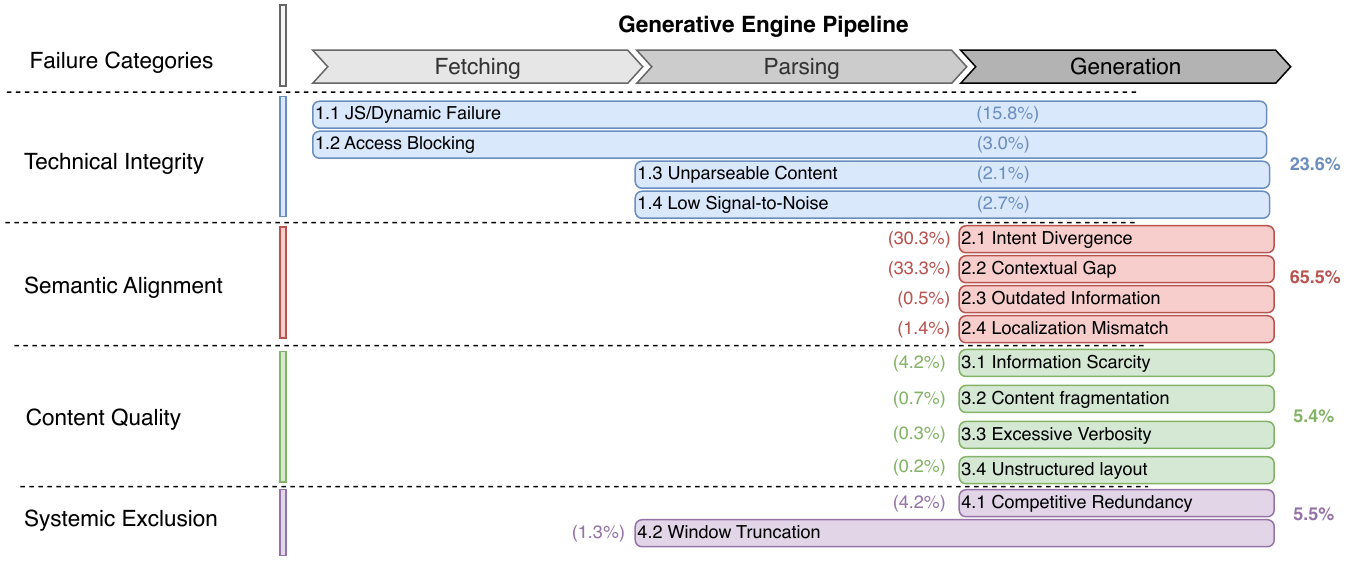}
    \caption{A taxonomy of citation failure modes in generative engines, spanning HTML fetching, parsing and answer generation stages.}
    \label{fig:taxnonomy}
\end{figure*}

\textbf{Contribution 2: Taxonomy of Citation Failures.} We introduce the first systematic taxonomy of citation failure modes spanning the generative engine pipeline: parsing-stage failures (malformed HTML, excessive noise), fetching/context failures (truncation, poor content ordering), and generation-stage failures (entity gaps, intent mismatch, competitor disadvantage). This taxonomy guides our tool library design and provides a foundation for future citation-focused optimization (Figure~\ref{fig:taxnonomy}).

\textbf{Contribution 3: Document-Centric Benchmark.} Existing benchmarks pair each document with a single query~\cite{aggarwal2024geo,wu2025generative}. In practice, content providers cannot anticipate exact queries; they need optimizations that generalize. We introduce MIMIQ (Multi-Intent Multi-Query), a document-centric benchmark associating each webpage with multiple queries spanning diverse intents, personas, and phrasings. This benchmark enables methods to optimize using a training query set and be evaluated on held-out queries, testing whether optimization produces genuinely more citable content rather than overfitting to specific formulations of a query.

\textbf{Empirical Takeaways.} Our experiments yield fresh insights into why webpages fail to achieve citation and what optimization can—and cannot—fix. 

\textbf{(1) Diagnosis beats generic rules while preserving content integrity (Section~\ref{ssec:over_perform})}.
AgentGEO achieves over 40\% relative improvement in citation rates across multiple generative engines. Crucially, it does so with minimal modification: baseline methods alter 25\% of original content on average, while our targeted repairs touch only 5\%. This suggests that citation failure is rarely a global content quality problem: most webpages need targeted fixes, not extensive rewriting.

\textbf{(2) Generic rules can harm long-tail content (Section~\ref{ssec:topic_analysis})}. We identify topic categories where generic optimization not only fails to improve citation but actively degrades it, whereas diagnostic optimization yields consistent gains. This divergence reflects a fundamental limitation: generic rules are derived from aggregate patterns, yet specialized topics and underrepresented domains deviate systematically. Diagnostic optimization, by conditioning on each webpage's specific failure mode, avoids this bias and generalizes more equitably across content types.

\textbf{(3) Not all citation failures are recoverable (Section~\ref{ssec:case_study}).} For certain webpages, even diagnostic optimization fails to improve citation. These webpages face disadvantages—dominant competitors—that no content-side modification can overcome. This finding has implications beyond optimization: if some content is systematically disadvantaged regardless of effort, citation mechanisms may amplify certain voices over others, and creator-side optimization alone cannot ensure equitable visibility.

\section{Related Work}

\subsection{Generative Engines}
\label{ssec:ge}
% LLMs has spurred their application across various domains, with GEs emerging as a notable implementation. 
GEs integrate LLMs to retrieve, synthesize, and generate responses directly based on web content utilizing retrieval-augmented generation (RAG) techniques~\cite{lewis2020retrieval, karpukhin2020dense}.
The workflow of a GE can be roughly divided into three stages: (1) retrieval, (2) content extraction (fetching), and (3) response generation. The retrieval stages identify and retrieve relevant web pages based on user queries. This process may employ traditional search engines~\cite{google2025aioverviews, willison2025anthropicbrave} or specialized neural retrieval models~\cite{karpukhin2020dense}. The content extraction stage focuses on parsing useful text from the retrieved pages, transforming unstructured web content into structured information suitable for LLM processing. Finally, the response generation stage leverages LLMs to produce coherent, contextually relevant answers derived from the extracted evidence.
Recent work further leverages LLM reasoning to iteratively plan and gather evidence for complex queries~\cite{jin2025search, mo2025conversational}. These developments board the research focus to the improvement of factual consistency and reliability in responses~\cite{salemi2024evaluating, niu2024ragtruth}.

% Beyond these foundational workflows, recent research has advanced more sophisticated GEs by incorporating the reasoning capabilities of LLMs to iteratively plan, reason, and gather evidence for complex queries~\cite{jin2025search, mo2025conversational}. These developments have broadened the research focus to encompass not only the integration of retrieval and generation but also the improvement of factual consistency and reliability in responses~\cite{salemi2024evaluating, niu2024ragtruth}.

\subsection{From SEO to GEO}
% Search engine optimization (SEO) has long been a pivotal area in information retrieval~\cite{davis2006search,yalccin2010search}. 
Traditional search engine optimization (SEO) improves webpage ranking through keyword optimization, backlink management, and site structuring ~\cite{davis2006search,yalccin2010search,killoran2013use}. As search shifts toward GEs, GEO has emerged to enhance the visibility of content within generated responses. 
Existing GEO research has explored two directions of textual modification: rewriting content with manually crafted heuristics~\cite{aggarwal2024geo,chen2025generative} or LLM-distilled preference rules~\cite{wu2025generative} to improve legitimate visibility, and crafting adversarial inputs such as prompt injection attacks~\cite{nestaas2024adversarial,pfrommer2024ranking,tang2025stealthrank} to manipulate GE behaviors. Despite their differing intents, both lines operate exclusively at the content-level textual modifications. They often neglect critical structural and systemic factors, such as HTML parsing mechanisms, context window constraints, and webpage structure, which significantly influence how content is ingested by the engine. In contrast, \name presents a holistic GEO framework that addresses these limitations by optimizing across multiple aspects of the webpage representation.

\subsection{GEO Benchmarks}

GEO-Bench~\cite{aggarwal2024geo} is one of the first GEO benchmarks, adopting a query-centric design that pairs each document with a single query and measures visibility gains relative to competing documents after optimization. E-commerce~\cite{wu2025generative} and Researchy-GEO~\cite{wu2025generative} follow the same setting but focus on specific domains. Yet this query-centric setup optimizes and evaluates on the same query, making it impossible to tell whether gains reflect genuinely more citable content or overfitting to that query's specific phrasing. CC-GSEO-Bench~\cite{chen2025ccgseobench} adopts a document-centric evaluation, associating each document with multiple queries; yet the number of queries per document is limited and inconsistent, preventing a comprehensive assessment of overall visibility. Moreover, none of these benchmarks evaluate raw HTML structure. Our MIMIQ benchmark (Section~\ref{sec:dataset}) addresses both gaps with a document-centric design, and explicit train/test query splits.

\section{Understanding Citation Failures}
\label{sec:understanding}

We adopt citation---the binary event of being included as a source---as our optimization target. To understand why documents fail to be cited, we construct a taxonomy of failure modes from contrastive cited–uncited document pairs.
% We focus on inline and pre-hoc attribution, the two paradigms used by major commercial systems (Perplexity~\cite{PerplexityAI2026}, ChatGPT search~\cite{OpenAI2024ChatGPTSearch}, Claude search~\cite{anthropic2026citations}); a detailed comparison of attribution paradigms is provided in Appendix~\ref{ssec:mechanisms}.

% Formally, for a query $q$ and a set of retrieved documents $\mathcal{D} = \{d_1, \ldots, d_n\}$, let $I_q \subseteq \{1, \ldots, n\}$ denote the indices of documents cited in the generated response. A document $d_i$ is \emph{cited} for query $q$ if $i \in I_q$. Across a query set $Q$, we define the \emph{citation rate}:
% \begin{equation}
%     \text{CR}(d_i) = \frac{1}{|Q|}\sum_{q \in Q} \mathbb{1}[i \in I_q]
% \end{equation}
% This metric captures a document's visibility---the probability it will be cited and thus have an opportunity to receive traffic.

% To understand why documents fail to achieve citation, we first examine how generative engines implement citation (\S\ref{ssec:mechanisms}), then present a taxonomy of failure modes (\S\ref{ssec:taxonomy}). These foundations inform our problem formulation (\S\ref{ssec:problem}) and the design of our diagnostic framework (Section~\ref{sec:method}).

\subsection{Taxonomy Construction}
\textbf{Data.} We construct 949 cited–uncited document pairs from GEO-Bench~\cite{aggarwal2024geo}, which spans 25 diverse domains (health, travel, technology, etc.). For each query where a retrieved document was not cited, we pair it with a cited competitor. This pairing isolates the marginal factor: both documents were retrieved for the same query, so the difference lies in why the engine preferred one over the other.

\textbf{Pipeline.} Taxonomy construction proceeded in three stages. (1) Failure pattern extraction: for each pair, an LLM-based diagnostic pipeline compared the two documents against the query and produced a failure diagnosis (prompts in Appendix~\ref{ssec:prompt_failure_taxonomy}). (2) Taxonomy design: three annotators (graduate students in NLP) collaboratively grouped recurring failure modes into categories and mapped each to the corresponding pipeline stage, producing the taxonomy in Figure 2. (3) Label validation: three annotators independently re-labeled a random sample of 100 pairs; reaching 84\% agreement with LLM labels (Cohen's $\kappa$ = 0.70), confirming the reliability of automated labeling at scale.

\subsection{The Citation Failure Taxonomy}
\label{ssec:taxonomy}

The taxonomy follows a sequential diagnostic procedure applied to each (query, cited document, uncited document) triple. Each stage corresponds to one failure category, and a document is assigned to the earliest category at which it fails. This design reflects the natural pipeline order—a document must be technically extractable before semantic relevance is assessable, and semantically relevant before content presentation matters.
Detailed explanation of each category is provided in Appendix~\ref{ssec:taxonomy_apx}.

% Why does a topically relevant webpage fail to be cited? To answer this systematically, we construct a diagnostic dataset from GEO-Bench~\cite{aggarwal2024geo}, which spans 10 diverse domains (health, travel, technology, etc.) and includes queries with varying intents. For each query where a retrieved webpage was not cited, we pair it with a cited competitor, yielding 949 contrastive pairs. This pairing strategy isolates the marginal factor: both webpage were retrieved for the same query, so the difference lies in why the engine preferred one over the other. By analyzing these pairs, we identify four primary failure dimensions mapped across the generative engine pipeline (Figure~\ref{fig:taxnonomy}). Detailed explanation of each category is provided in Appendix~\ref{ssec:taxonomy}.
% \ruoxi{briefly talk where are 949 pairs from? why they are are reasonable starting point to extract the taxonomy, like are they comprehensive? are they representative? etc. overall, we should rationalize our starting point}

\textbf{S1: Technical Integrity (23.6\%).} \textit{Can the document be ingested at all?} Failures occur at the \emph{fetching} stage (JavaScript rendering failures, access blocking) or the \emph{parsing} stage (unparseable content, low signal-to-noise from boilerplate). A webpage with critical information buried in a JavaScript widget or overwhelmed by navigation chrome may never surface to the generator.

\textbf{S2: Semantic Alignment (65.5\%).} \textit{Does the extracted content contain information that addresses the query?} This includes intent divergence, contextual gaps (missing specific entities or terminology), outdated information, and localization mismatch. We operationalize this check by extracting the answer evidence from the cited document and verifying whether equivalent information exists in the uncited document. For instance, a query about U.S. tax filing deadlines returns a page on UK self-assessment dates—topically related but for the wrong jurisdiction.

% The webpage's content does not match what the query requires. This includes \emph{intent divergence} (informational content for a transactional query), \emph{contextual gaps} (missing specific entities or terminology), \emph{outdated information}, and \emph{localization mismatch} (e.g., UK regulations for a US query). These are generation-stage failures: content reaches the model but is judged insufficiently relevant.

\textbf{S3: Content Quality (5.4\%).} \textit{Is the relevant information presented in a form that supports citation?} 
For documents containing query-relevant information yet remaining uncited, the failure lies in how the information is presented:  information scarcity, content fragmentation, excessive verbosity, or unstructured layout. For instance, a document answers a "how does X work" query with three pages of background and a single sentence on the actual mechanism buried near the end—the answer is present, but not in a form the engine surfaces.

% The webpage addresses the right topic but presents information poorly. Manifestations include \emph{information scarcity} (too shallow to cite), \emph{content fragmentation} (disconnected snippets resisting synthesis), \emph{excessive verbosity} (key facts diluted by filler), and \emph{unstructured layout} (dense prose where tables would aid extraction). These are generation-stage failures where better-presented competitors win.

\textbf{S4: Systemic Exclusion (5.5\%).}\textit{The document passes S1–S3 yet remains uncited; the failure is attributed to factors external to the document itself.} The engine's fixed context window may truncate relevant content, or the model may preferentially cite a higher-authority source with the same facts (e.g., Wikipedia). While document-side mitigations can partially help (e.g., content relocation for truncation), the root cause is architectural; we empirically characterize this category in Section ~\ref{ssec:case_study}.

\begin{figure*}
    \centering
    \includegraphics[width=1\textwidth]{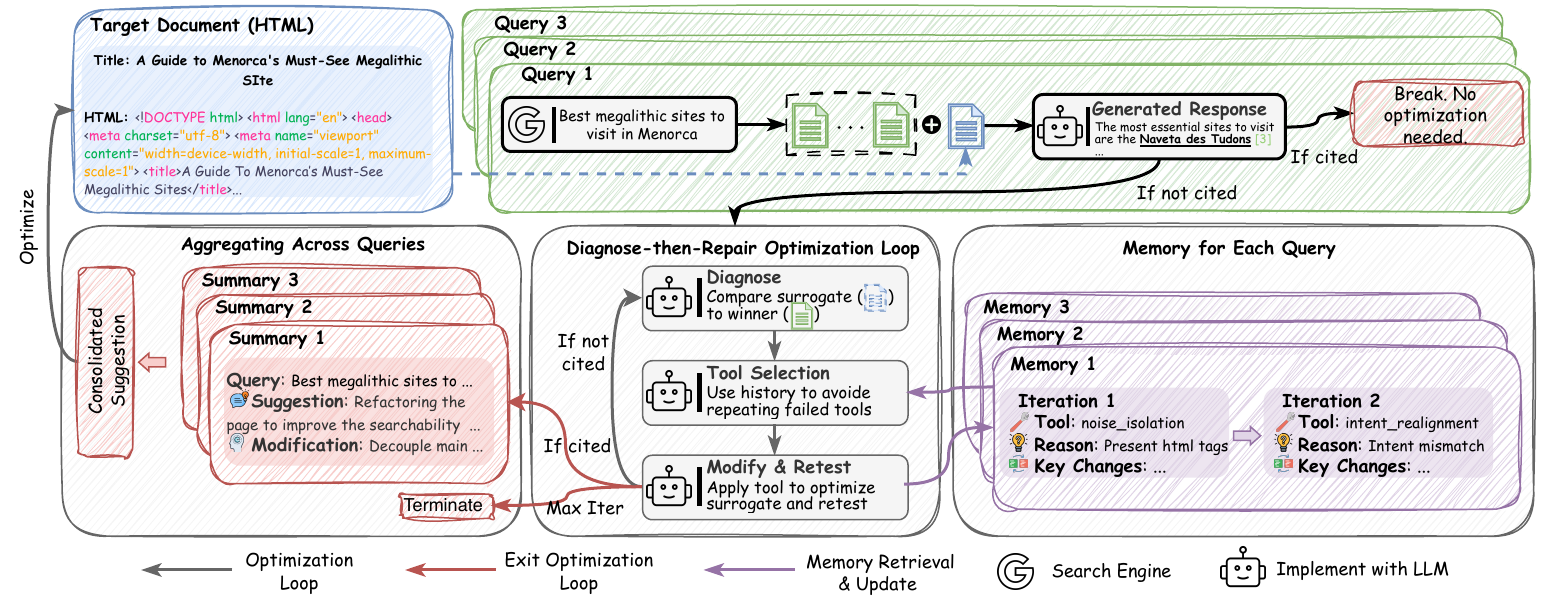}
    \caption{\textbf{Optimization process of \name.} 
    \name first checks citation status against competitor documents. For uncited ones, it runs a "Diagnose-then-Repair" loop: (1) an LLM diagnoses why the document was not cited; (2) guided by the diagnosis and a query-specific memory of prior iterations, a tool is selected to edit a surrogate copy of the webpage; (3) the surrogate is retested against the same competitors. The loop continues until citation succeeds or a limit is hit. Successful edits are aggregated into a single suggestion applied to the original page, and the procedure iterates over additional training queries for further optimization.}
    \label{fig:workflow}
\end{figure*}

\section{AgentGEO: A Diagnostic Approach to GEO}

\paragraph{Problem Setup.} Let $d$ denote a target webpage and $\mathcal{Q}$ a distribution of user queries relevant to $d$. For query $q$, let $\mathcal{V}(q, d, \mathcal{G}) \in \{0,1\}$ indicate whether webpage $d$ is cited by generative engine $\mathcal{G}$. We seek an optimized webpage $d^*$ that maximizes expected citation across $\mathcal{Q}$ while preserving fidelity to the original:
\begin{equation*}
d^* = \arg\max_{d'} \; \mathbb{E}_{q \sim \mathcal{Q}} \left[ \mathcal{V}(q, d', \mathcal{G}) \right] \quad \text{s.t.} \quad \text{Sim}(d, d') \geq \tau
\end{equation*}
In practice, we optimize using training queries $Q_{\text{tr}}$ and evaluate on held-out queries $Q_{\text{te}}$. 
% Our taxonomy (\S\ref{ssec:taxonomy}) suggests this problem admits efficient solutions: most citation failures stem from presentation issues rather than fundamental content gaps, so targeted repairs---guided by diagnosis---should suffice.

\textbf{Overview.} AgentGEO follows a three-stage pipeline (Figure~\ref{fig:workflow}). First, for each training query, a \emph{diagnose-then-repair} cycle identifies why the target webpage fails to be cited and applies targeted fixes, guided by the taxonomy and a memory module that tracks prior attempts. Second, a \emph{batch aggregation} stage merges query-specific suggestions into query-invariant modifications, preventing overfitting to individual queries. Third, a \emph{localized editing} mechanism applies modifications at the chunk level, preserving semantic integrity.

\subsection{Diagnose-then-Repair Loop}
\label{ssec:instance_level_opt}

For each training query $q$, AgentGEO executes an iterative cycle that transforms the target webpage $d$ until it achieves citation or the iteration limit is reached.

\textbf{Failure Diagnosis.}
\label{ssec:failure_dia}
We simulate the GE pipeline: retrieve candidate webpage, generate a response, and record which sources are cited. If the target webpage is not cited, diagnosis begins. To identify the failure reason, the system compares $d$ against the highest-ranked cited competitor and characterizes the vulnerability, i.e., why the competitor was cited instead. These vulnerabilities are classified according to our taxonomy of failure modes (Section~\ref{ssec:taxonomy}).
% which spans fetching, parsing, and generation stages.

\textbf{Tool Selection with Memory.}
Given the diagnosed vulnerability, \name selects a repair tool from a library of specialized interventions spanning HTML repair, content reordering, entity enrichment, and information density improvements (full list in Appendix~\ref{appdix:tools}). To avoid repeating ineffective repairs, we maintain a query-specific memory that tracks prior attempts. At each iteration $t$, the agent selects a tool based on the current vulnerability and memory. Memory is initialized empty.
% \begin{equation}
% % \omega^{(t)} = \pi(v_q^{(t)}, \mathcal{M}_q, \Omega)
% \end{equation}
% where $\pi$ is the selection policy, $v_q^{(t)}$ is the diagnosed vulnerability, and $\Omega$ is the tool library. 

\textbf{Iterative Refinement.}
% The selected tool is applied to produce a modified webpage $d^{(t)}$. If citation is achieved, the system extracts a suggestion summarizing the successful edits—specifically, which content locations were modified and how. These suggestions feed into batch aggregation described next. If citation is not achieved, memory is updated with the failed attempt, and the system re-diagnoses against competitors. This loop continues until citation succeeds or the iteration limit is reached.
Each iteration applies the selected tool to produce a modified webpage $d^{(t)}$ and queries the GE to test whether $d^{(t)}$ is now cited. On success, the system records a \emph{suggestion}—the edit locations and the operations applied—which is later consumed by batch aggregation (Section~\ref{ssec:batch_wise_aggr}). On failure, the system appends the failed attempt to memory and re-runs diagnosis against the current competitors, which may select a different tool in the next iteration. The loop terminates when citation succeeds or the iteration budget is exhausted.

\subsection{Aggregating Across Queries}
\label{ssec:batch_wise_aggr}

% To generalize beyond individual training queries, \name aggregates suggestions across a batch rather than applying them one by one. Specifically, suggestions that succeeded for multiple queries are prioritized, as cross-query consistency indicates robustness. When multiple suggestions target the same location with conflicting edits, the most frequently successful edit is retained. This ensures that only query-invariant modifications survive aggregation. The filtered suggestions are then applied to produce the optimized document $d^*$, subject to the faithfulness constrain $\text{Sim}(d, d^*) \geq \tau$.
To achieve generalization across training queries, AgentGEO performs webpage updates at the batch level. For each query in a batch, we run the diagnose-then-repair loop (Section~\ref{ssec:failure_dia}) on an isolated, temporary copy of the current HTML. Each run yields at most one candidate edit, consisting of the diagnosed failure type, a target structural chunk index, and a tool-produced HTML fragment to update that chunk. To keep edits comparable across queries, we fix the chunk partition at the start of the batch and map all candidate
edits into this frozen index space, even if the temporary DOM changes during iterative retries.
We aggregate candidate edits by target chunk and, for each chunk, select at most one edit to apply to the
shared webpage. Our default diagnosis-aware selector ranks candidates by failure severity and a confidence
score, resolving conflicts by retaining the highest-ranked proposal. (Other conflict-resolution strategies
include heuristic voting within a chunk or an LLM-based arbiter.) The selected per-chunk edits are applied in
a single pass to synthesize the batch update, mitigating query-specific overfitting and promoting query-
invariant improvements. We do not enforce an explicit similarity threshold at aggregation time; fidelity is
primarily maintained through localized chunk-level editing (Section~\ref{ssec:chunking}).

\subsection{Localized Editing for faithfulness}
\label{ssec:chunking}

To satisfy the semantic preservation constraint, AgentGEO applies modifications at the chunk level rather than rewriting entire webpages. This serves two purposes: it preserves faithfulness by restricting edits to targeted locations, and it avoids the quality degradation LLMs exhibit when processing long webpages~\cite{liu2024lost}.
The webpage is partitioned into structure-aware chunks using HTML tags, where each chunk represents a coherent unit (e.g., a paragraph, list, or metadata block). Each suggestion from the aggregation stage specifies the content location it targets; we map these locations to the corresponding chunks. Modifications are applied only to matched chunks; the remaining content and overall HTML structure remain unchanged.

\section{Document-Centric Benchmarking}
\label{sec:dataset}

% Existing benchmarks for GEO typically evaluate document visibility under single-query settings. However, real-world retrieval environments are characterized by diverse user intents and phrasing variability. To address this gap, we introduce \textbf{MIMIQ} (\textbf{M}ulti-\textbf{I}ntent \textbf{M}ult\textbf{I}-\textbf{Q}uery), a document-centric dataset designed to evaluate the effectiveness of optimization methods across a distribution of queries.

Existing GEO benchmarks pair each document with a single query and evaluate optimization on that same query~\cite{aggarwal2024geo}. This setup risks conflating genuine improvements with query-specific overfitting: an optimization that succeeds for one query may fail for another expressing the same intent.

We introduce \benchmark (\textbf{M}ulti-\textbf{I}ntent \textbf{M}ult\textbf{I}-\textbf{Q}uery), a document-centric benchmark for evaluating GEO methods. We term it \textit{document-centric} because it treats the document as the unit of optimization and evaluation, mirroring the content creator's setting: they control their document but cannot anticipate the exact queries users will issue. For each document, \benchmark provides multiple queries spanning diverse intents, personas, and phrasings, with explicit train/test splits so that optimizations are evaluated on held-out queries. A detailed comparison to existing benchmarks is provided in Table~\ref{tab:comparison} (Appendix~\ref{app:dataset_details}).

% Methods optimize using a training query set and are evaluated on held-out queries unseen during optimization. This protocol tests whether an optimization produces genuinely more citable content—or merely overfits to specific query formulations. 
% Table~\ref{tab:comparison} highlights the distinct features of MIMIQ compared to concurrent benchmarks.

\subsection{Construction Pipeline}
\textbf{Webpage Sourcing.}
We construct a candidate webpage pool from two sources, using each only to obtain webpages; all queries in MIMIQ are subsequently generated from scratch (next paragraph). First, we collect URLs from GEO-Bench~\cite{aggarwal2024geo} query--document pairs and fetch their original HTML. Second, to diversify domain coverage, we issue seed queries to ChatNoir~\cite{bevendorff2018elastic} over the ClueWeb22 index~\cite{overwijk2022clueweb22} and collect the returned pages. Candidates are deduplicated by URL and stratified by topic and content length to prevent skew toward any single domain or short pages. After manual review, we retain 204 webpages spanning 20+ topics with content lengths from a few hundred to over 10{,}000 words. Sampling heuristics and distribution statistics are provided in Appendix~\ref{app:dataset_details}.

% To simulate a realistic search landscape, we
% We use LLM to generate a dense set of queries for each target webpage. To generate queries that reflect realistic search diversity, we design a \emph{structured} generation framework spanning three dimensions: intent, persona, and phrasing.
% Detailed prompt templates and filtering heuristics are provided in Appendix~\ref{app:dataset_details}.

\textbf{Query Generation.} For each webpage, we generate queries through a three-dimensional framework spanning \textit{intent}, \textit{persona}, and \textit{phrasing}. As a preprocessing step, we extract anchor keywords from the page title, as it best reflects the page's topic, organizing them into three categories: core keywords, latent semantic synonyms, and long-tail phrases.
% We do not use the full content, as it tends to surface peripheral terms that drift queries away from the page's core topic—for instance, a page about \textit{safe road trips} may yield ``maps'' as a keyword. 
We then generate queries by combining each keyword with values sampled along three dimensions: (1) \textbf{intent}, drawn from the standard taxonomy of Navigational, Informational, Transactional, and Commercial searches~\cite{broder2002taxonomy,rose2004understanding}, with implausible combinations filtered out (e.g., navigational intent for a recipe page); (2) \textbf{persona}, synthesized to capture heterogeneous user backgrounds (e.g., ``Domain Expert'' vs. ``Casual Browser''); and (3) \textbf{phrasing}, generating multiple paraphrases per intent--persona combination to capture lexical and syntactic variation. A verifier step removes hallucinations and ensures each query is logically answerable by the target webpage. This yields 60 queries per webpage, partitioned into 20 training and 40 test queries, for a total of 12,240 queries. Prompt templates are provided in Appendix~\ref{ssec:query_generation_apx}.

% \paragraph{Content-Anchored Keyword Extraction.}
% Directly utilizing entire web page content for query generation often introduces noise. For example, extracting keywords from a page about "safe road trips" may yield irrelevant terms like "maps".
% To maintain relevance, we extract keywords exclusively from the page \texttt{title} metadata, which typically encapsulates the core topic concisely. We instruct the LLM to derive three categories of terms: (1) \textit{Core Keywords} for topical relevance, (2) \textit{Synonyms/Latent semantic indexing terms} for semantic expansion, and (3) \textit{Long-tail Phrases} to capture specific user needs.

% \paragraph{Persona and Intent Modeling.}
% To mimic heterogeneous user behaviors, we incorporate personality simulation and intent classification. 
% We synthesize diverse user personas (e.g., ``Domain Expert'' vs. ``Casual Browser'') and map them to standard search intents (Navigational, Informational, Transactional, Commercial)~\cite{broder2002taxonomy,rose2004understanding}. For each webpage, we filter out implausible intents (e.g., removing navigational intent for a pure recipe page) to ensure the generated queries are contextually valid.

% \paragraph{Quality Control.}
% The raw generated queries undergo an additional deduplication and relevance filtering process. We employ a verifier model to eliminate hallucinations, ensuring that every query in the final set is logically answerable by the target webpage.

\subsection{Dataset Variants}
Beyond the standard MIMIQ, we build two variants targeting specific optimizer capabilities. Each comprises 50 webpages with 60 queries per page, and follows the same train/test split protocol.

% The standard MIMIQ dataset comprises 204 webpages sampled from the ClueWeb22 index. For each webpage, we curate a stratified set of 60 queries, split into 20 for training and 40 for testing, resulting in a total of 12,240 queries. Beyond the standard setting, we introduce two challenging variants to probe specific optimizer capabilities:
\begin{itemize}[leftmargin=*,topsep=0pt]
    \item \textbf{MIMIQ-OOD (Out-of-Distribution):} To evaluate generalization to unseen user behaviors, we cluster queries by persona and assign disjoint persona sets to training and test queries. An optimizer that overfits to training-persona phrasings will degrade on held-out personas.
    
    \item \textbf{MIMIQ-HTML (Structural Robustness):} We curate webpages with complex or noisy DOM structures—cookie notices, navigation residue, JavaScript artifacts, login walls, etc. This variant evaluates whether an optimizer can repair upstream parsing failures, not just rewrite clean text.

    % In practice, citation failures often originate in the upstream fetching and parsing phases. MIMIQ-HTML subset identified as having complex or irregular DOM structures. This variant specifically evaluates the optimizer's ability to parse and modify raw HTML.
    % under technically challenging conditions.
\end{itemize}
% Each variant comprises 50 webpages with 60 queries per page, and maintaining the same train/test split as the standard dataset.

\section{Experiments}
\label{sec:exp}
% In this section, we conduct extensive experiments to evaluate the effectiveness of \name. 

\subsection{Experimental Settings}
\textbf{Baselines.}
We consider three types of baselines. The Vanilla baseline is the original generative engine without any optimization. GEO-Bench~\cite{aggarwal2024geo} proposes eight methods that optimize website content using LLMs. AutoGEO~\cite{wu2025generative} automatically summarizes preference rules from a large number of webpage pairs and optimizes website content accordingly.
% We test all the methods on generative engines built with state-of-the-art LLMs, including GPT (gpt-4.1-mini, gpt-4.1), and Claude (claude-haiku-4-5-20251001). gpt-4.1 is more advanced then the other two llms.

\textbf{Metrics.}
We evaluate model performance along three dimensions. \textbf{Visibility} measures the extent to which the optimized webpage is surfaced by the GE in its response: we report citation rate (CR) — the fraction of test queries on which the optimized webpage is cited — together with Word, Pos, and Wordpos from GEO-Bench~\cite{aggarwal2024geo} to characterize lexical and positional contribution. \textbf{Faithfulness} measures the retention of original information after optimization, computed as TF-IDF, Embedding, and Jaccard similarity between the original and optimized content. We further evaluate the \textbf{utility} of generated content, including content quality and citation quality in Appendix~\ref{sec:exp_apx}. All results are reported as percentages (\%). Full metric definitions are in Appendix~\ref{sec:metric_definitions_apx}.

\textbf{Generative Engine Setup.}
We construct a controlled simulation using standard retrieval and generation components with explicit citation instructions. This enables systematic diagnosis, reproducible evaluation, and controlled ablations infeasible with black-box commercial systems.
\textbf{(1) Retrieval.} 
We use ChatNoir~\cite{bevendorff2018elastic} over the ClueWeb22~\cite{overwijk2022clueweb22} corpus (10B webpages) to retrieve the top-5 candidates per query. Since our optimized webpages are not in ChatNoir's index, end-to-end native retrieval is infeasible; we therefore append the target webpage as the final candidate. This isolates the GEO question (\emph{does optimized content get cited more?}) from the orthogonal SEO question (whether the page is retrieved). The placement is also fair by construction: optimized and non-optimized versions of the target are served from the \emph{same} fixed position, so any position-induced bias affects both conditions identically and cancels in the comparison.
\textbf{(2) Content extraction.} We parse HTML with BeautifulSoup~\cite{richardson2025beautifulsoup} to retain semantic structure.
\textbf{(3) Generation.} We use GPT-4.1-mini as the default generator under two citation paradigms: inline attribution via in-context generation~\cite{gao2023enabling}, and pre-hoc attribution via attribute-first-then-generate~\cite{slobodkin2024attribute}. These dominate commercial systems (Perplexity~\cite{PerplexityAI2026}, ChatGPT Search~\cite{OpenAI2024ChatGPTSearch}, Claude search~\cite{anthropic2026citations}).

% Specifically, we use text-embedding-3-small from openai to generate document embeddings.
\begin{table*}[t] 
\small
\centering
\caption{Performance comparison of \name against baselines across different citation methods. \textbf{Bold} and \underline{underline} indicate the best and second-best
results, respectively. "-" indicates the metric is not applicable to the method.}
\label{tab:citation_methods}
\resizebox{\textwidth}{!}{
\setlength{\tabcolsep}{2pt}
\begin{tabular}{lcccccccccccccc} 
\toprule
& \multicolumn{7}{c}{\textbf{In-context}} & \multicolumn{7}{c}{\textbf{Attribute first then generate}} \\ 
\cmidrule(lr){2-8} \cmidrule(lr){9-15}
& \multicolumn{1}{c}{Citation} & \multicolumn{3}{c}{Contribution} & \multicolumn{3}{c}{Faithfulness} & \multicolumn{1}{c}{Citation} & \multicolumn{3}{c}{Contribution} & \multicolumn{3}{c}{Faithfulness} \\
\cmidrule(lr){2-2} \cmidrule(lr){3-5} \cmidrule(lr){6-8} \cmidrule(lr){9-9} \cmidrule(lr){10-12} \cmidrule(lr){13-15}
\textbf{Method} & \textbf{CR} & \textbf{Word} & \textbf{Pos} & \textbf{Wordpos} & \textbf{TF-IDF} & \textbf{Embed} & \textbf{Jaccard} & \textbf{CR} & \textbf{Word} & \textbf{Pos} & \textbf{Wordpos} & \textbf{TF-IDF} & \textbf{Embed} & \textbf{Jaccard} \\
\midrule
Vanilla & 56.58 & 16.28 & 16.28 & 16.39 & - & - & - & 60.20 & 23.69 & 28.72 & 23.57 & - & - & - \\ 
\midrule
Technical Terms & 55.84 & 14.78 & 14.73 & 14.60 & 73.32 & 85.64 & 21.37 & 57.33 & 22.25 & 27.87 & 22.03 & 74.39 & 86.98 & 21.98 \\ 
Cite Sources & 60.52 & 17.01 & 16.96 & 16.89 & 92.76 & 93.62 & 65.62 & 59.21 & 24.04 & 29.02 & 23.79 & 92.33 & 92.95 & 64.60 \\ 
Keyword Stuffing & 62.44 & 17.83 & 17.74 & 17.82 & 82.85 & 88.95 & 30.04 & 65.62 & 28.09 & 33.39 & 27.81 & 82.16 & 86.92 & 28.39 \\ 
Unique Words & 57.76 & 15.33 & 15.37 & 15.19 & 86.00 & 91.58 & 32.72 & 57.42 & 23.43 & 28.45 & 23.23 & 86.11 & 91.58 & 33.30 \\
Authoritative & 59.14 & 17.45 & 17.75 & 17.57 & 80.42 & 86.38 & 25.15 & 63.46 & 26.86 & 31.85 & 26.63 & 80.85 & 86.35 & 26.02 \\
Easy-to-Understand & \underline{71.11} & \underline{19.73} & \underline{19.84} & \underline{19.75} & 82.17 & 86.68 & 30.26 & 65.34 & 28.44 & \underline{33.62} & 27.98 & 81.99 & 86.93 & 30.41 \\ 
Statistics Addition & 60.52 & 17.41 & 17.29 & 17.26 & \underline{95.52} & \textbf{94.19} & 69.31 & 59.46 & 24.15 & 29.39 & 23.93 & 95.51 & \underline{93.75} & 69.38 \\  
Quotation Addition & 59.00 & 16.28 & 16.16 & 16.19 & \textbf{96.78} & \underline{93.92} & \underline{74.85} & 59.99 & 23.66 & 28.88 & 23.48 & \underline{96.57} & \textbf{94.29} & \underline{74.78} \\ 
Fluency Optimization & 60.60 & 16.52 & 16.63 & 16.48 & 92.40 & 92.72 & 48.98 & 62.22 & 25.44 & 30.95 & 25.25 & 92.20 & 92.83 & 49.21 \\ 
\midrule
AutoGEO & 68.80 & 17.87 & 18.53 & 17.65 & 67.49 & 76.99 & 17.97 & \underline{65.97} & \underline{33.96} & \textbf{34.33} & \textbf{32.88} & 74.55 & 78.52 & 19.77 \\
\midrule
AgentGEO (ours) & \textbf{79.52} & \textbf{20.29} & \textbf{21.08} & \textbf{20.28} & 94.23 & 93.20 & \textbf{82.40} & \textbf{70.00} & \textbf{34.18} & 31.57 & \underline{30.92} & \textbf{97.44} & 92.77 & \textbf{82.66} \\  
\bottomrule 
\end{tabular}
}
\end{table*}

%% -----------------------------------------------------

\begin{table}[t]\centering
\caption{Performance comparison across different LLM-based generative engines. "-" indicates the metric is not applicable to the method.}
\label{tab:llm}
\small
\setlength{\tabcolsep}{4pt}
\renewcommand{\arraystretch}{1.05}
\resizebox{\linewidth}{!}{
\begin{tabular}{l ccc c ccc c ccc}
\toprule
& \multicolumn{3}{c}{\textbf{GPT-4.1-mini}} & & \multicolumn{3}{c}{\textbf{GPT-4.1}} & & \multicolumn{3}{c}{\textbf{Claude-haiku-4.5}} \\
\cmidrule(lr){2-4} \cmidrule(lr){6-8} \cmidrule(lr){10-12}
\textbf{Metric} & \textbf{Vanilla} & \textbf{AutoGEO} & \textbf{\name (ours)} & & \textbf{Vanilla} & \textbf{AutoGEO} & \textbf{\name (ours)} & & \textbf{Vanilla} & \textbf{AutoGEO} & \textbf{\name (ours)} \\
\midrule
CR              & 56.58 & 68.80 & \textbf{79.52} & & 76.97 & 82.43 & \textbf{84.74} & & 42.40 & 46.80 & \textbf{54.80} \\
\cmidrule(lr){1-4} \cmidrule(lr){6-8} \cmidrule(lr){10-12}
Word            & 16.28 & 17.87 & \textbf{20.29} & & 28.55    & 31.67    & \textbf{32.77}             & & 21.80 & 24.86 & \textbf{26.94} \\
Pos             & 16.28 & 18.53 & \textbf{21.08} & & 28.70    & 31.84    & \textbf{32.53}             & & 21.54 & 24.81 & \textbf{27.57} \\
Wordpos         & 16.39 & 17.65 & \textbf{20.28} & & 28.68    & 31.84    & \textbf{32.35}             & & 21.44 & 24.66 & \textbf{26.10} \\
\cmidrule(lr){1-4} \cmidrule(lr){6-8} \cmidrule(lr){10-12}
Precision       & 48.63 & 56.79 & \textbf{58.43} & & 82.32    & \textbf{89.25}     & 82.91             & & 89.42 & 88.79 & \textbf{90.33} \\
Recall          & 64.62 & \textbf{68.99} & 67.10  & & 88.02    & \textbf{88.22}    & 88.07             & & \textbf{96.16} & 95.51 & 96.02 \\
Clarity         & 61.23 & \textbf{66.08} & 62.88  & & 74.62    & \textbf{74.88}    & 72.95             & & 73.99 & \textbf{75.74} & 74.99 \\
Insightfulness  & 40.20 & \textbf{45.25} & 42.68  & & 64.52    & \textbf{64.58}    & 61.56             & & 57.47 & \textbf{61.05} & 60.76 \\
\cmidrule(lr){1-4} \cmidrule(lr){6-8} \cmidrule(lr){10-12}
TF-IDF          & --    & 67.49 & \textbf{94.23} & & --    & 73.56 & \textbf{94.29} & & --    & 74.27 & \textbf{95.26} \\
Embed           & --    & 76.99 & \textbf{93.20} & & --    & 67.97 & \textbf{93.85} & & --    & 79.98 & \textbf{88.76} \\
Jaccard         & --    & 17.97 & \textbf{82.40} & & --    & 19.56 & \textbf{69.26} & & --    & 22.28 & \textbf{74.58} \\
\bottomrule
\end{tabular}
}
\end{table}
%% -----------------------------------------------------

\begin{table*}[t]
\centering
\caption{Performance comparison in OOD and HTML compatibility settings.}
\label{tab:ood_html}
\resizebox{\textwidth}{!}{ 
\begin{tabular}{l*{7}{c}c*{7}{c}}
\toprule
& \multicolumn{7}{c}{\textbf{MIMIQ-OOD}} & & \multicolumn{7}{c}{\textbf{MIMIQ-HTML}} \\
\cmidrule(lr){2-8} \cmidrule(lr){10-16}
& Citation & \multicolumn{3}{c}{Contribution} & \multicolumn{3}{c}{Faithfulness} & & Citation & \multicolumn{3}{c}{Contribution} & \multicolumn{3}{c}{Faithfulness} \\
\cmidrule(lr){2-2} \cmidrule(lr){3-5} \cmidrule(lr){6-8} \cmidrule(lr){10-10} \cmidrule(lr){11-13} \cmidrule(lr){14-16}
\textbf{Method} & \textbf{CR} & \textbf{Word} & \textbf{Pos} & \textbf{Wordpos} & \textbf{TF-IDF} & \textbf{Embed} & \textbf{Jaccard} & & \textbf{CR} & \textbf{Word} & \textbf{Pos} & \textbf{Wordpos} & \textbf{TF-IDF} & \textbf{Embed} & \textbf{Jaccard} \\
\midrule
Vanilla & 57.14 & \textbf{25.19} & \textbf{25.01} & \textbf{24.94} & - & - & - & & 38.14 & 18.20 & 18.41 & 18.20 & - & - & - \\
AutoGEO & \underline{64.18} & 20.19 & 20.86 & 20.66 & \underline{73.09} & \underline{79.43} & \underline{19.48} & & \underline{45.96} & \underline{23.79} & \underline{24.12} & \underline{23.81} & \underline{65.77} & \underline{81.51} & \underline{17.39} \\
\midrule
\name (ours) & \textbf{70.62} & \underline{24.57} & \underline{24.78} & \underline{24.42} & \textbf{96.43} & \textbf{92.27} & \textbf{83.28} & & \textbf{67.19} & \textbf{32.10} & \textbf{32.43} & \textbf{32.15} & \textbf{82.20} & \textbf{95.04} & \textbf{63.33} \\
\bottomrule
\end{tabular}
}
\end{table*}
%% -----------------------------------------------------

\subsection{Overall GEO Performance}
\label{ssec:over_perform}

\textbf{Performance across different citation methods.}
We compare \name with two baseline citation methods: in-context generation~\cite{gao2023enabling} and attribute-first-then-generate~\cite{slobodkin2024attribute}. In in-context generation, the generative engine is directly prompted to produce answers with citations based on the optimized webpage. In attribute-first-then-generate, it first extracts citation-related attributes from the webpage and then generates answers from them.
Table~\ref{tab:citation_methods} presents the results. \name consistently outperforms all baselines across both citation methods and generative engines, demonstrating its effectiveness and robustness. In in-context generation, \name achieves a citation rate of 79.52\%, outperforming the best baseline, AutoGEO, by 10.72. In attribute-first-then-generate, it attains 70.00\%, surpassing AutoGEO by 4.03. \name also excels in contribution and faithfulness metrics, improving citation rates while maintaining high-quality, faithful content.

\textbf{Performance across different LLM backbones.}
To assess whether \name's advantages generalize across generators, we evaluate 50 webpages sampled from the standard dataset using three LLMs: GPT-4.1-mini, GPT-4.1, and Claude-haiku-4.5. As shown in Table~\ref{tab:llm}, \name consistently achieves the highest citation rate, outperforming Vanilla by 22.94, 7.77, and 12.40 percentage points, respectively. Claude generally produces higher-quality responses across all methods, with higher Precision, Recall, Clarity, and Insightfulness than GPT-4.1-mini, suggesting stronger answer synthesis capabilities regardless of optimization. Within each generator, all methods achieve comparable utility scores, indicating that \name's citation gains do not compromise response quality. Moreover, \name consistently achieves higher faithfulness than AutoGEO across all generators. On GPT-4.1-mini, its Jaccard similarity reaches 82.40 compared with AutoGEO's 17.97, demonstrating that diagnosis-guided optimization better preserves original content than rule-based rewriting.

\textbf{Performance under OOD and HTML-compatible settings.}
Table~\ref{tab:ood_html} reports the results. On MIMIQ-OOD, \name attains a 70.62\% citation rate, exceeding AutoGEO by 6.44 points, and ranks first on all three faithfulness metrics. We attribute this margin to information preservation: \name modifies only 3.6\% of each page's content (TF-IDF), targeting the text responsible for citation failure while leaving the rest intact to support other queries; AutoGEO instead regenerates the entire webpage as a single LLM rewrite, modifying 26.9\% of TF-IDF content—roughly 7$\times$ more—and discarding information unrelated to the rewriting query but useful for OOD queries at evaluation time. The advantage is amplified on MIMIQ-HTML, where \name reaches a 67.19\% citation rate—a 21.2-point margin over AutoGEO—and achieves the best score on every contribution and faithfulness metric. This gap confirms that HTML structure materially affects citation outcomes and cannot be handled by treating webpages as plain text; \name addresses this through dedicated HTML-aware editing tools that operate on the page's structural elements rather than discarding them, while AutoGEO's text-level rewrite loses this information.

% Table~\ref{tab:ood_html} presents a performance comparison across out-of-distribution (OOD) and HTML compatibility settings. In the MIMIQ-OOD scenario, \name achieves a citation rate of 70.62\%, outperforming AutoGEO by 6.44\%. This performance gap is largely attributable to the differing transformation strategies: while AutoGEO often suffers from detrimental information attrition—incurring a 43\% average reduction in webpage length—\name maintains webpage length through granular augmentation, preserving the informational density required to satisfy diverse OOD queries. Furthermore, whereas AutoGEO's complete rewriting process is prone to semantic drift and query-induced bias, \name ensures higher robustness by safeguarding original informational nodes. Notably, the \textit{Easy-to-Understand} method achieves competitive citation performance by utilizing extreme distillation and structural reorganization to isolate key informational anchors, demonstrating that enhanced clarity can effectively offset its 53\% reduction in average length.
% In the MIMIQ-HTML setting, \name attains a citation rate of 67.19\%, significantly higher than baselines, demonstrating its effectiveness in optimizing HTML-compatible webpages. \name also excels in contribution and faithfulness metrics, validating its capability to enhance webpage quality while ensuring HTML compatibility.

%% -----------------------------------------------------
\subsection{Experiments on a Commercial Engine}

To assess practical applicability, we evaluate \name on a commercial search engine. We query ChatGPT in web-search mode—the same mode used in its commercial deployment—across a range of queries, and collect two types of webpages: those cited in the response, and those retrieved but not cited. We optimize the latter on our simulated generative engine, then re-query ChatGPT with the optimized pages included in the candidate set. Experiments are conducted with GPT-5-mini over 160 test queries. Since optimized pages are not in ChatGPT's index, we follow~\cite{nestaas2024adversarial} and host them at public URLs, explicitly instructing ChatGPT to fetch them. This preserves the engine's native fetching and parsing pipeline. Both optimized and original versions of each webpage are evaluated under the identical instruction protocol. 
\name achieves a citation rate of 48\%, compared with 39\% for Vanilla—a 23\% relative improvement. While it is slightly below Vanilla on Word and Position, it surpasses on WordPos.
This confirms that our optimizations can transfer across a commercial engine. 

% To assess practical applicability, we evaluate \name on a commercial search engine. We test webpages optimized in our simulated pipeline (GPT-4.1-mini) against GPT-5 with web search -- the same mode used in ChatGPT's commercial deployment. This cross-mode setup also tests the transferability of optimization across different LLMs.
% Since we cannot control ChatGPT's index and thus cannot guarantee retrieval of a newly hosted page, we follow \cite{nestaas2024adversarial} and host both optimized and candidate webpages as public URLs, explicitly instructing ChatGPT to read them.
% This preserves the engine's native fetching, parsing, and generation.
% Results show that \name achieves a citation rate of 96\%, compared with 86\% for Vanilla, a 71\% reduction in failure rate. This confirms that optimizations from our pipeline transfer across model generations to a commercial engine. The high Vanilla baseline stems from bypassing retrieval. In organic search, where pages must first be discovered, the room for citation-focused optimization should grow rather than shrink.

\begin{table}[h]
\centering
\scriptsize
\setlength{\tabcolsep}{4pt}
\renewcommand{\arraystretch}{0.95}
\caption{Experiments on commercial engine}
\begin{tabular}{lcccc}
\toprule
\textbf{Method} & \textbf{Citation Rate} & \textbf{Word} & \textbf{Position} & \textbf{Wordpos} \\
\midrule
% Vanilla  & 86.00 & 25.53 & 27.40 & 26.12 \\
Vanilla  & 39.34 & \textbf{14.19} & \textbf{11.24} & 14.03\\
AgentGEO & \textbf{48.15} & 14.11 & 9.61 & \textbf{14.09} \\
\bottomrule
\end{tabular}
\end{table}

\subsection{Experiments in an End-to-End Retrieval Setting}

Our previous experiments include the target webpage as the last candidate provided to the generator. While this allows us to evaluate citation performance when the target is included, it does not capture how webpage modifications affect whether the target is retrieved. To examine this effect, we construct a search index containing the target webpage and competing documents, and rank them using BM25~\cite{robertson2009probabilistic}, a widely adopted lexical ranking function. Only the top-$K$ retrieved documents are passed to the generator, allowing webpage modifications to affect both retrieval and subsequent citation.
For each target, we construct a competitor pool of size $N$ and pass the top-$K$ retrieved candidates to the generator. We measure visibility, citation conditional on visibility, and end-to-end citation, where $P(\text{cited}) = P(\text{visible}) \cdot P(\text{cited} \mid \text{visible})$. As shown in Table~\ref{tab:bm25_results}, \name improves end-to-end citation over Vanilla by 14.5--16.2 percentage points across all four settings. The gains reflect both increased retrieval visibility and higher conditional citation rates. For example, at $N=50$ and $K=10$, visibility increases from 51.7\% to 67.4\%, while conditional citation rises from 85.2\% to 89.4\%, improving end-to-end citation from 43.6\% to 59.8\%. These results demonstrate that \name remains effective when target exposure depends on organic retrieval rather than guaranteed inclusion.

\begin{table}[t]
\centering
\caption{Retrieval visibility and citation performance under
BM25 retrieval. $N$ is the competitor pool size, and $K$ is the number of retrieved documents passed to the generator.}
\label{tab:bm25_results}
\small
\setlength{\tabcolsep}{3.5pt}
\begin{tabular}{@{}ccrrrrrr@{}}
\toprule
\multirow{2}{*}{$N$} &
\multirow{2}{*}{$K$} &
\multicolumn{2}{c}{Visible (\%)} &
\multicolumn{2}{c}{Cite$|$Vis (\%)} &
\multicolumn{2}{c}{Cited (\%)} \\
\cmidrule(lr){3-4}
\cmidrule(lr){5-6}
\cmidrule(lr){7-8}
& & Vanilla & AgentGEO
  & Vanilla & AgentGEO
  & Vanilla & AgentGEO \\
\midrule
30 & 10 & 68.2 & \textbf{80.2} & 84.8 & \textbf{91.6}
   & 57.3 & \textbf{73.0} \\
50 & 10 & 51.7 & \textbf{67.4} & 85.2 & \textbf{89.4}
   & 43.6 & \textbf{59.8} \\
30 &  5 & 47.2 & \textbf{60.6} & 91.1 & \textbf{94.7}
   & 42.6 & \textbf{57.1} \\
50 &  5 & 33.5 & \textbf{47.7} & 90.0 & \textbf{95.6}
   & 29.9 & \textbf{45.3} \\
\bottomrule
\end{tabular}
\end{table}

\subsection{Evaluation of Different Topics}
\label{ssec:topic_analysis}
\begin{wrapfigure}[18]{r}{0.5\linewidth}
    \centering
    \includegraphics[width=\linewidth]{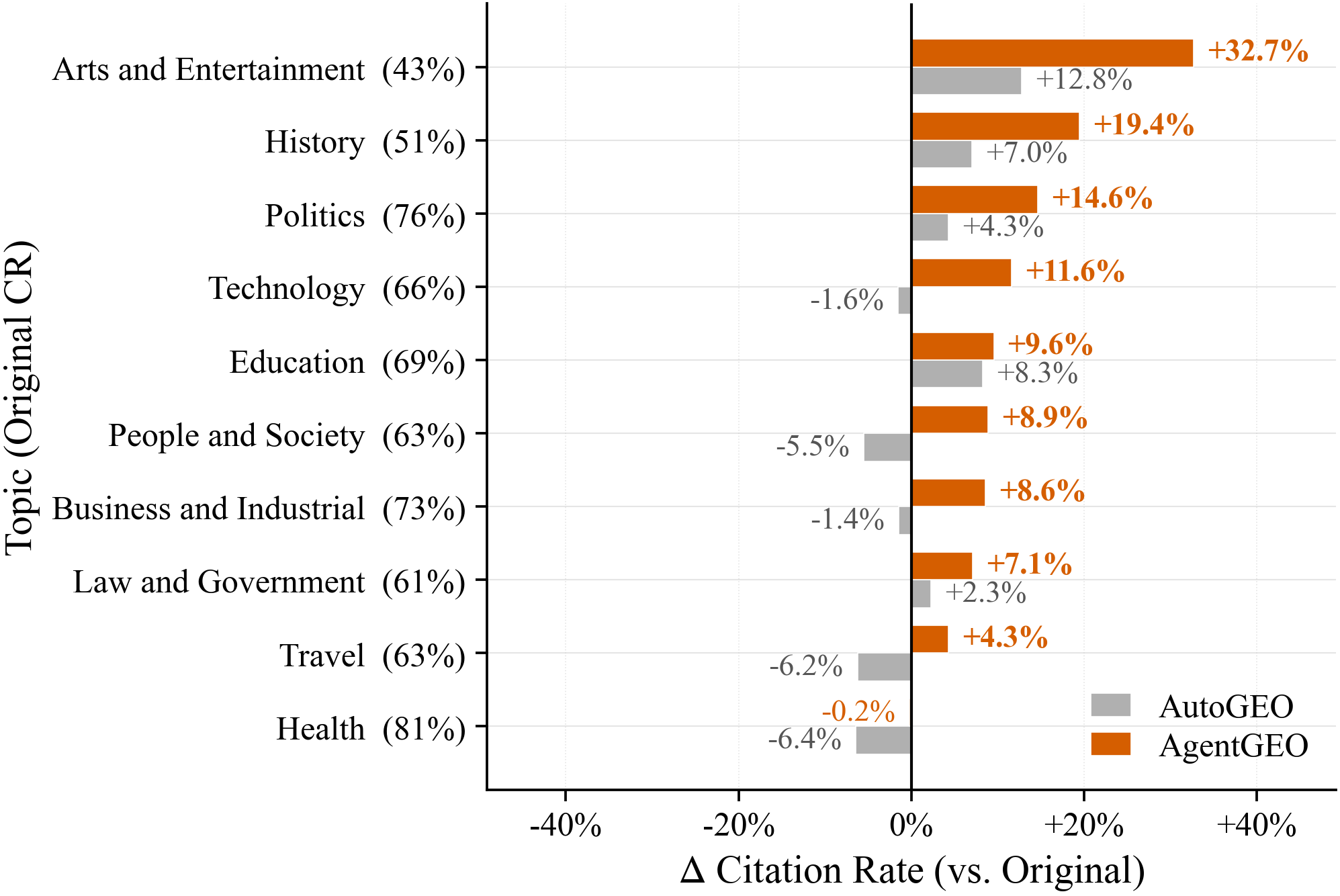} 
    \caption{Citation rate change over vanilla baseline across different topics. \name improves 9 of 10 topics while AutoGEO degrades in 5 of 10, particularly those with already-high baseline rates.}
    \label{fig:topic_analysis}
\end{wrapfigure}
Figure~\ref{fig:topic_analysis} breaks down citation rate improvement over the vanilla baseline by topic (arts, history, etc.). \name yields positive improvements across nearly all topics, with the largest gains on arts and entertainment, history, and politics. AutoGEO, in contrast, drops below the vanilla baseline on 5 out of 10 topics—particularly those with already high original citation rates—indicating that a single rewriting rule applied uniformly cannot accommodate topic-specific failure modes and risks degrading content that did not need fixing. \name avoids this regression because each webpage is diagnosed individually, and edits to one document do not constrain edits to another. The exception is the health topic, where the vanilla citation rate is already near 80\%, leaving little room for improvement.

\subsection{Failure Optimization Case Study}
\label{ssec:case_study}
To probe the boundaries of \name, we examine the failure cases from our main optimization run: although the overall citation rate rises from 57.0\% to 83.7\% on training queries, 163 queries still fail to elicit a citation of their target webpage. Inspecting these cases reveals a consistent pattern: the optimization layer is not the bottleneck. The pages have, in most cases, been successfully edited to address the failure mode \name diagnosed for them, yet the GE still elects to cite a different source. The failures lie not in our pipeline's ability to act on a diagnosis, but in the GE's preference for sources whose advantage is not reducible to page content.
Consider a university course page on machine learning targeted by the query \emph{``best online machine learning courses''}. \name expands the page with detailed course descriptions, prerequisites, and learning outcomes, addressing the diagnosed coverage gap. The GE nevertheless cites Coursera in its place, presumably reflecting source-level priors—platform reputation, domain authority—that operate independently of page content. Such priors form a hard ceiling for any purely content-based optimization, and call for deeper investigation into the systemic factors underlying GE citation mechanisms.

\section{Conclusion}

This paper presents a diagnostic approach to Generative Engine Optimization that shifts focus from generic content rewriting to understanding why individual webpages fail to be cited. We introduce a taxonomy of citation failure modes, \name—an agentic system that diagnoses failures and applies targeted repairs—and a webpage-centric benchmark for comprehensive evaluation. 
% Our experiments demonstrate that \name achieves over 40\% relative improvement in citation rates while modifying only 5\% of content. 
% Beyond performance gains, our analysis reveals important insights: generic optimization strategies can inadvertently harm long-tail content, and some webpages face structural barriers that no optimization can overcome. These findings highlight the need for more nuanced, diagnostic approaches to GEO and raise important considerations for equitable visibility in AI-mediated information access.
Future work could focus on validating \name against commercial engines and improving scalability through distillation and efficient diagnosis strategies.

\begin{ack}
We acknowledge support from the National Science Foundation through grants IIS-2312794, IIS-2313130, OAC-2239622, CNS-2424127, OAC-2613761, and the Amazon--Virginia Tech Initiative for Efficient and Robust Machine Learning.
\end{ack}

\bibliography{main}
\bibliographystyle{unsrt}

%%%%%%%%%%%%%%%%%%%%%%%%%%%%%%%%%%%%%%%%%%%%%%%%%%%%%%%%%%%%%%%%%%%%%%%%%%%%%%%
%%%%%%%%%%%%%%%%%%%%%%%%%%%%%%%%%%%%%%%%%%%%%%%%%%%%%%%%%%%%%%%%%%%%%%%%%%%%%%%
% APPENDIX
%%%%%%%%%%%%%%%%%%%%%%%%%%%%%%%%%%%%%%%%%%%%%%%%%%%%%%%%%%%%%%%%%%%%%%%%%%%%%%%
%%%%%%%%%%%%%%%%%%%%%%%%%%%%%%%%%%%%%%%%%%%%%%%%%%%%%%%%%%%%%%%%%%%%%%%%%%%%%%%
\newpage
\appendix

\section*{Appendix Overview}

This appendix provides additional details on benchmark construction,
evaluation, experiments, and the implementation of AgentGEO.

\begin{center}
\small
\renewcommand{\arraystretch}{1.15}
\begin{tabular}{@{}lp{0.70\linewidth}@{}}
\toprule
\textbf{Section} & \textbf{Contents} \\
\midrule
\hyperref[app:dataset_details]{A}. Benchmark Construction
    & Data collection, sampling, and benchmark variants \\
\hyperref[ssec:mechanisms]{B}. Citation Mechanisms
    & Citation paradigms in generative engines \\
\hyperref[sec:metric_definitions_apx]{C}. Evaluation Metrics
    & Citation, faithfulness, and utility metrics \\
\hyperref[app:cost]{D}. Cost Analysis
    & LLM calls and optimization cost \\
\hyperref[sec:exp_apx]{E}. Additional Experiments
    & Cross-model transfer, robustness, and ablations \\
\hyperref[ssec:taxonomy_apx]{F}. Citation Failure Taxonomy
    & Failure-mode definitions and examples \\
\hyperref[appdix:tools]{G}. Optimization Tools
    & Repair operators and tool-selection policy \\
\hyperref[app:prompt]{H}. Prompt Templates
    & Prompts used throughout the framework \\
\bottomrule
\end{tabular}
\end{center}

\section*{Broader Impacts}

This research introduces \name, a framework for Generative Engine Optimization (GEO). While GEO empowers content providers to remain visible in an evolving search landscape, it also introduces several potential risks that warrant ethical consideration:
\begin{itemize}[leftmargin=*,itemsep=0.5em,topsep=0.5em]
    \item  Information Integrity and Manipulation: The ability to systematically optimize content for LLM citations could be exploited to prioritize biased, low-quality, or even misleading information. If malicious actors use GEO to ensure their content is cited more frequently than authoritative sources, the reliability of generative engines as a primary information gatekeeper could be severely compromised.
    
    \item  Degradation of Human-Centric Content: Similar to the historical "SEO-spam" in traditional search, a widespread adoption of GEO may incentivize creators to write for "engines" rather than humans. This could lead to a web ecosystem filled with "LLM-friendly" but redundant or stylistically unnatural content, ultimately degrading the quality of information available to human readers.
\end{itemize}
% To mitigate these risks, we emphasize that GEO research should proceed in tandem with the development of robust, fact-checking generative engines. By introducing \benchmark (MIMIQ), we encourage the community to evaluate GEO not just through citation gains, but through the lens of document faithfulness and utility, ensuring that optimization serves to enhance, rather than distort, the global information ecosystem.

% \onecolumn

\section{Benchmark Construction}
\label{app:dataset_details}

We construct the MIMIQ benchmarks using real-world web data. Specifically, we leverage GEO-Bench and ChatNoir to build a candidate pool of web pages, from which we sample and filter to curate the final dataset. GEO-Bench provides query-document pairs, each consisting of a query and five candidate documents. We use the associated URLs to retrieve the original HTML pages.

To diversify the data sources, we supplement the dataset using ChatNoir's search~\cite{bevendorff2018elastic} over the ClueWeb22 index~\cite{overwijk2022clueweb22}. Specifically, we issue 40 diverse queries spanning 10 broad domains (science, health, technology, history, business, education, environment, law, sports, and arts), each returning up to 50 results. Retrieved documents are deduplicated against the GEO-Bench candidates by URL.

\paragraph{MIMIQ.}
We categorize each document into four length groups by words count: Short ($<$2K), Medium (2K--5K), Long (5K--10K), and Very Long ($>$10K). Based on length group and topic, we apply a priority-based sampling strategy: (1) at least 100 documents must belong to the Medium, Long, or Very Long groups to prevent domination by short web pages; (2) at least 20 distinct topics must be represented, prioritizing 15 high-importance topics (e.g., science, health, technology, history); and (3) remaining slots are distributed evenly across topics to maintain balance. This yields 210 candidates, which after manual review produce the final MIMIQ benchmark of 204 documents.

\begin{figure}[htbp]
    \centering
    \begin{subfigure}[b]{0.5\linewidth}
        \centering
        \includegraphics[width=\linewidth]{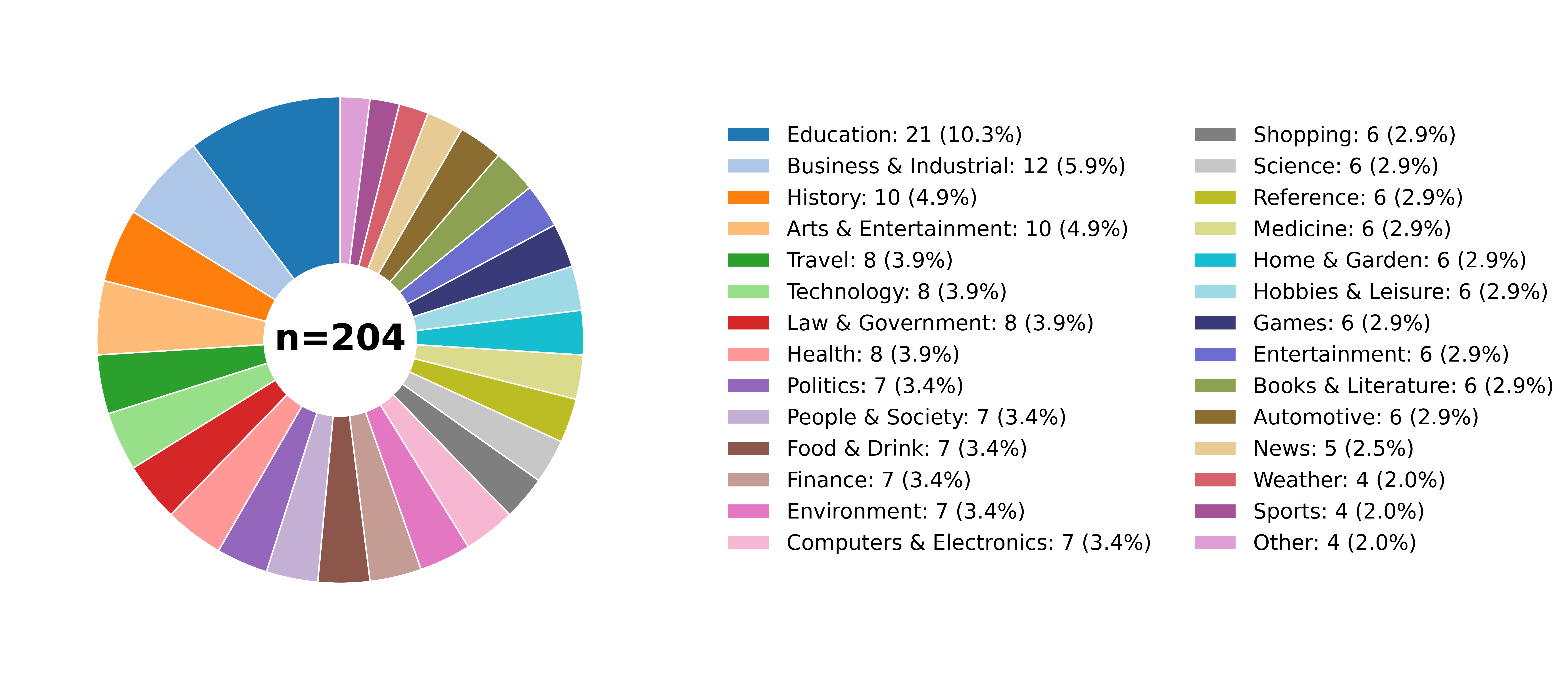}
        \caption{Topic Distribution}
        \label{fig:topic_dist}
    \end{subfigure}
    \begin{subfigure}[b]{0.42\linewidth}
        \centering
        \includegraphics[width=\linewidth]{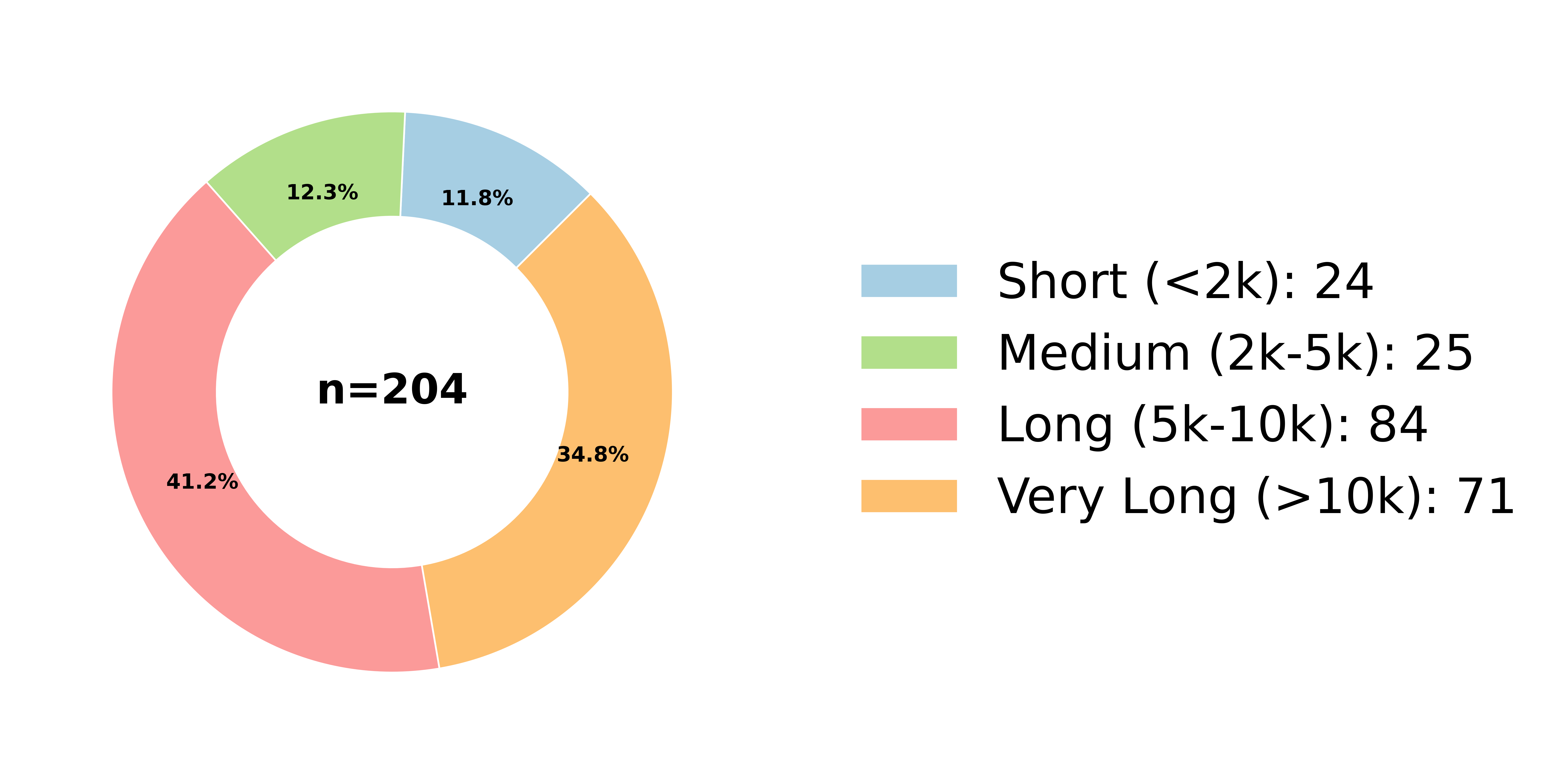}
        \caption{Content Length Distribution}
        \label{fig:length_dist}
    \end{subfigure}
    \caption{Distribution of topics and content lengths in the MIMIQ benchmark.}
    \label{fig:distribution}
\end{figure}

Figure~\ref{fig:distribution} presents the distribution of topics and content lengths within the MIMIQ benchmark. As illustrated in Figure~\ref{fig:topic_dist}, the benchmark spans a wide range of subjects, including technology, health, finance, education, and entertainment, ensuring comprehensive coverage of various domains. Figure~\ref{fig:length_dist} shows that webpage lengths vary significantly, ranging from brief articles of a few hundred words to extensive reports exceeding 10,000 words. This diversity in both thematic content and webpage size enhances the benchmark's utility for evaluating information retrieval and citation systems.

\paragraph{MIMIQ-OOD.}
MIMIQ-OOD is constructed by randomly sampling 50 documents from MIMIQ. The key difference lies in the train/test query split. Recall that we simulate diverse search intents by generating queries from different user personas (cf. Section~\ref{sec:dataset}); in MIMIQ-OOD, we assign disjoint sets of personas to generate training and test queries, enabling evaluation of generalization to unseen query distributions.

\paragraph{MIMIQ-HTML.}

From the same candidate pool, we curate a subset of 50 documents that exhibit common HTML parsing artifacts, including cookie notices, navigation residue, footer elements, JavaScript code residue, empty or 404 pages, encoding issues, boilerplate text (e.g., ``click here''), and login walls. This subset enables targeted evaluation of systems' robustness to noisy, imperfectly parsed web content.

\begin{table}[t]
\centering
\footnotesize
\renewcommand{\arraystretch}{1.0}
\caption{Comparison of \benchmark and existing GEO benchmarks. Our dataset combines a document-centric paradigm with a holistic evaluation scope and supports explicit train/test query splits.}
\label{tab:comparison}
\begin{tabular}{lcccc}
\toprule
\textbf{Dataset} & \textbf{Doc-centric} & \textbf{HTML} & \textbf{Train/Test Split} &\textbf{Queries/doc}\\
\midrule
GEO-Bench~\citep{aggarwal2024geo}/E-commerce~\citep{wu2025generative}/Researchy-GEO~\citep{wu2025generative}         & \xmark & \xmark & \cmark & 1 \\
% E-commerce~\citep{wu2025generative}        & \xmark & \xmark & \cmark \\
% Researchy-GEO~\citep{wu2025generative}     & \xmark & \xmark & \cmark \\
CC-GSEO-Bench~\citep{chen2025ccgseobench}  & \cmark & \xmark & \xmark & 2-10\\
\midrule
\textbf{\benchmark (Ours)} & \cmark & \cmark & \cmark & 60 \\
\bottomrule
\end{tabular}
\end{table}

\section{Citation Mechanisms in Generative Engines}
\label{ssec:mechanisms}

How a generative engine attributes its response to sources determines where citation failures can occur. We identify the following three paradigms.

\textbf{Inline Attribution.} Reference markers (e.g., ``[1]'') are embedded directly within the generated sequence, providing fine-grained provenance at the sentence or clause level. The model decides \emph{during generation} which sources support each claim~\cite{gao2023enabling}. Systems like Perplexity and ChatGPT's search mode employ this approach~\cite{PerplexityAI2026,OpenAI2024ChatGPTSearch}.

\textbf{Pre-hoc Attribution.} The system first selects passages from retrieved documents, then generates a response grounded exclusively in these selections. Because generation operates over pre-selected evidence, the source context serves as an inherent anchor---the model cannot easily hallucinate beyond what was extracted~\cite{slobodkin2024attribute}. However, relevant content not selected upstream will not be cited. System like Claude's search mode employ this approach~\cite{anthropic2026citations}.

\textbf{Post-hoc Attribution.} Generation and verification are decoupled: the model first produces a response, then a separate module retrospectively identifies supporting sources~\citep{bohnet2022attributed, gao2023enabling}. While flexible, this paradigm risks hallucination propagation---errors introduced during generation may persist if the verifier incorrectly validates unsupported claims.

Given their prevalence in commercial systems and tighter grounding guarantees, we focus on inline and pre-hoc attribution. These paradigms expose different failure surfaces: inline attribution can fail when the model overlooks a relevant source during generation, while pre-hoc attribution can fail earlier, during passage selection. Our taxonomy accounts for failures across both paradigms.

\section{Metric Definitions}
\label{sec:metric_definitions_apx}
We provide the definition of all metrics used in our experiments. We organize them along three dimensions: visibility, faithfulness, and utility.

\subsection{Visibility}

Let $\mathcal{Q}_{\text{te}}$ denote the held-out test query set for a target webpage $d$, and let $r_q$ denote the response generated by the GE for query $q$ with citations to retrieved documents.

\textbf{Citation Rate (CR).} The fraction of test queries on which the target webpage is cited at least once:
\begin{equation}
\text{CR}(d) = \frac{1}{|\mathcal{Q}_{\text{te}}|} \sum_{q \in \mathcal{Q}_{\text{te}}} \mathbb{1}[\,d \in \text{cited}(r_q)\,].
\end{equation}
This is our primary metric. CR captures the binary event of being cited, directly reflecting whether a webpage receives any referral opportunity from the GE.

\textbf{Word, Pos, Wordpos.} Following GEO-Bench~\cite{aggarwal2024geo}, these metrics assess the lexical and positional share that the cited webpage contributes to the response, conditional on being cited:
\begin{itemize}[leftmargin=*,topsep=0pt]
    \item \textbf{Word} measures the fraction of words in the response attributable to $d$, weighted by lexical overlap between cited spans and $d$'s content.
    \item \textbf{Pos} measures the positional prominence of $d$'s contribution, with citations appearing earlier in the response receiving higher weight.
    \item \textbf{Wordpos} combines lexical and positional share into a single score.
\end{itemize}
Whereas CR captures whether $d$ is cited at all, Word/Pos/Wordpos quantify how much of the response is attributable to $d$ given citation. We treat CR as primary and report Word/Pos/Wordpos as complementary. We refer readers to~\cite{aggarwal2024geo} for the precise formulations.

\subsection{Faithfulness}
Let $d$ denote the original webpage and $d'$ the optimized version. Faithfulness metrics quantify the preservation of original information after optimization.

\textbf{TF-IDF Similarity.} Cosine similarity between TF-IDF weighted term vectors of $d$ and $d'$:
\begin{equation}
\text{TF-IDF}(d, d') = \cos(\mathbf{v}^{\text{tfidf}}_d, \mathbf{v}^{\text{tfidf}}_{d'}).
\end{equation}
This captures preservation of term-level content with weighting by term importance, penalizing the loss or alteration of distinctive vocabulary.

\textbf{Embedding Similarity.} Cosine similarity between dense semantic embeddings of $d$ and $d'$, computed using OpenAI's \texttt{text-embedding-3-small}:
\begin{equation}
\text{Embed}(d, d') = \cos(\mathbf{e}_d, \mathbf{e}_{d'}).
\end{equation}
This captures semantic preservation: even if surface words change, content with similar meaning yields high similarity.

\textbf{Jaccard Similarity.} Strict lexical overlap between unique word sets:
\begin{equation}
\text{Jaccard}(d, d') = \frac{|W(d) \cap W(d')|}{|W(d) \cup W(d')|},
\end{equation}
where $W(\cdot)$ denotes the set of unique tokens after standard preprocessing. Jaccard is the strictest faithfulness metric: any word added or removed reduces the score, making it a sensitive indicator of how much surface content has been altered.

The three metrics jointly characterize faithfulness from term-weighted, semantic, and strict lexical perspectives.

\subsection{Utility}

Utility metrics evaluate whether the optimization improves the end product---the GE-generated response---rather than just citation visibility. We decompose utility into content quality and citation quality.

\textbf{Content Quality.} Following DeepResearchGym~\cite{coelho2025deepresearchgym}, we evaluate generated responses along two axes:
\begin{itemize}[leftmargin=*,topsep=0pt]
    \item \textbf{Relevance} measures how well the response addresses the query. We report \textbf{Precision} (the fraction of response content that is relevant to the query) and \textbf{Recall} (the fraction of query-relevant information present in the response).
    \item \textbf{Quality} measures the readability and substantiveness of the response. We report \textbf{Clarity} (coherence and readability, judged by an LLM evaluator) and \textbf{Insightfulness} (depth and substantiveness, judged by an LLM evaluator).
\end{itemize}

\textbf{Citation Quality.} Citation quality measures whether the citations to the optimized webpage genuinely support the generated claims. Following Gao et al.~\cite{gao2023enabling}, we report:
\begin{itemize}[leftmargin=*,topsep=0pt]
    \item \textbf{Citation Recall}: the fraction of generated statements that are fully supported by their cited passages.
    \item \textbf{Citation Precision}: the fraction of non-removable citations, i.e., citations whose removal would leave the corresponding statement unsupported.
\end{itemize}
Together, citation recall measures whether claims have adequate evidentiary support, and citation precision measures whether each citation is necessary rather than redundant. Entailment for both metrics is judged by an LLM evaluator following the protocol of~\cite{gao2023enabling}.

\section{Cost Analysis}
\label{app:cost}

Each diagnose-then-repair iteration involves four LLM calls (diagnosis, tool selection, repair, and citation re-verification), compared to two calls for both GEO-Bench and AutoGEO (rewriting, and re-verification). With 20 training queries per document and an average of two iterations per query, AgentGEO requires approximately 160 LLM calls per document, costing around \$0.49 using GPT-4.1-mini.
This overhead is justified by AgentGEO's substantially higher citation rate and content preservation, particularly given that the cost remains modest in absolute terms.

\section{Additional Experiments}
\label{sec:exp_apx}
This section presents additional experiments on generalization (\ref{app:deep_research}–\ref{app:cross_model_transfer}), content and optimization analysis (\ref{ssec:length_analysis}–\ref{sec:scaling_analysis}), and red-teaming and ablation studies (\ref{app:red_teaming}–\ref{app:ablation}).

\subsection{Performance on Deep Research Systems}
\label{app:deep_research}
Deep research systems are analogous to generative engines: both retrieve web content and synthesize cited responses. Unlike single-pass GEs, however, deep research systems iteratively identify knowledge gaps and issue fine-grained follow-up searches before producing a final answer. This iterative behavior makes them a meaningful alternative evaluation environment for testing whether GEO methods transfer beyond single-pass settings.
We evaluate on a deep research system implemented with the open-source \texttt{qx-labs/agents-deep-research} framework~\cite{qxlabs_deep_research}.
Table~\ref{tab:deep_research} reports the results. AgentGEO improves citation rate from 78.00\% to 84.47\%, an absolute gain of 6.47 points over Vanilla, and substantially improves all contribution metrics, with Wordpos increasing from 27.65 to 34.31, a 24.1\% relative gain. Notably, AutoGEO \emph{decreases} citation rate to 69.00\% in this setting, despite achieving comparable contribution scores to Vanilla. This pattern echoes our finding in Section~\ref{ssec:topic_analysis}: generic rewriting rules can hurt visibility when the target system has its own preferences, whereas diagnostic optimization remains robust. Content quality metrics are comparable across all methods, indicating that AgentGEO's citation gains do not come at the cost of response quality. These results suggest that AgentGEO's diagnostic approach generalizes beyond standard GE pipelines to iterative retrieval-synthesis systems.

\begin{table}[t]
\centering
\small
\caption{Performance on a deep research system.}
\label{tab:deep_research}
\begin{tabular}{lcccccccc}
\toprule
& Citation & \multicolumn{3}{c}{Contribution} & \multicolumn{4}{c}{Content Quality} \\
\cmidrule(lr){2-2} \cmidrule(lr){3-5} \cmidrule(lr){6-9}
Method & \textbf{CR} & \textbf{Word} & \textbf{Pos} & \textbf{Wordpos} & \textbf{Precision} & \textbf{Recall} & \textbf{Clarity} & \textbf{Insight} \\
\midrule
Vanilla            & 78.00 & 28.11 & 27.39 & 27.65 & \textbf{11.07} & 86.43 & 68.72 & 47.40 \\
AutoGEO            & 69.00 & 27.82 & 28.26 & 28.42 & \phantom{0}9.22 & \textbf{86.84} & \textbf{70.80} & \textbf{50.45} \\
\midrule
\textbf{AgentGEO (ours)} & \textbf{84.47} & \textbf{34.60} & \textbf{34.00} & \textbf{34.31} & \phantom{0}9.86 & 85.92 & 68.15 & 48.59 \\
\bottomrule
\end{tabular}
\end{table}

\subsection{Cross-Citation Mechanism Robustness Analysis}
\label{sec:cross_method}
To assess the generalization capability of \name across different citation mechanisms, we conducted a cross-method robustness experiment on a subset of 50 webpages. We enforced a distribution shift between optimization and evaluation: \name was optimized using \textit{Attribute-First} feedback but evaluated under the \textit{In-Context} generation paradigm. This setup strictly tests whether structural repairs translate across distinct engine architectures. As shown in Table~\ref{tab:cross_method_robustness}, \name demonstrates strong transferability, improving the Citation Rate by \textbf{14.31} (from 69.98\% to 84.29\%) and boosting Contribution metrics by over 9\%. Coupled with high faithfulness scores (TF-IDF: 94.91\%, Embedding: 90.32\%), these results confirm that \name addresses fundamental content deficiencies rather than overfitting to specific engine artifacts.

\begin{table*}[t] 
\centering
\caption{Cross-Citation Mechanism Robustness Evaluation.}
\label{tab:cross_method_robustness}
\small
\setlength{\tabcolsep}{8pt} 
\begin{tabular}{l c ccc ccc}
\toprule
 & Citation & \multicolumn{3}{c}{Contribution} & \multicolumn{3}{c}{Faithfulness} \\
\cmidrule(lr){2-2} \cmidrule(lr){3-5} \cmidrule(lr){6-8}
Method & CR & Word & Pos & Wordpos & TF-IDF & Embed & Jaccard \\
\midrule
Vanilla & 69.98 & 26.09 & 25.98 & 25.93 & -- & -- & -- \\
\name & \textbf{84.29} & \textbf{35.18} & \textbf{34.89} & \textbf{34.54} & 94.91 & 90.32 & 75.36 \\
\midrule
\textit{Improv.} & \textit{+14.31} & \textit{+9.09} & \textit{+8.91} & \textit{+8.61} & -- & -- & -- \\
\bottomrule
\end{tabular}
\end{table*}

\subsection{Cross-Model Transfer of AgentGEO Optimizations}
\label{app:cross_model_transfer}

To evaluate whether webpages optimized by \name transfer to different LLM families, we optimize webpages using GPT-4.1-mini and evaluate the original and optimized versions with two other generators: DeepSeek-V4-Flash and Kimi-K2.6. As shown in Table~\ref{tab:cross_model_transfer}, the citation rate increases from 69.20 to 76.74 on DeepSeek-V4-Flash and from 71.47 to 78.74 on Kimi-K2.6. The Word, Pos, and WordPos metrics also improve on both generators. These results indicate that the improvements obtained using GPT-4.1-mini also transfer to the two evaluated model families.

\begin{table}[t]
\centering
\caption{Cross-model transfer of AgentGEO optimizations.
Webpages are optimized using GPT-4.1-mini and evaluated
with two other generators.}
\label{tab:cross_model_transfer}
\footnotesize
\setlength{\tabcolsep}{3pt}
\begin{tabular}{@{}llcccc@{}}
\toprule
\multirow{2}{*}{Model} &
\multirow{2}{*}{Method} &
\multicolumn{1}{c}{Citation} &
\multicolumn{3}{c}{Contribution} \\
\cmidrule(lr){3-3}
\cmidrule(lr){4-6}
& & CR & Word & Pos & WordPos \\
\midrule
\multirow{3}{*}{DeepSeek-V4-Flash}
& Original
    & 69.20 & 25.44 & 24.91 & 24.97 \\
& \name
    & \textbf{76.74} & \textbf{30.38}
    & \textbf{29.48} & \textbf{29.68} \\
\cmidrule(lr){2-6}
& \textit{Improv.}
    & \textit{+7.54} & \textit{+4.94}
    & \textit{+4.57} & \textit{+4.71} \\
\midrule
\multirow{3}{*}{Kimi-K2.6}
& Original
    & 71.47 & 21.89 & 22.19 & 21.85 \\
& \name
    & \textbf{78.74} & \textbf{26.68}
    & \textbf{27.20} & \textbf{26.59} \\
\cmidrule(lr){2-6}
& \textit{Improv.}
    & \textit{+7.27} & \textit{+4.79}
    & \textit{+5.01} & \textit{+4.74} \\
\bottomrule
\end{tabular}
\end{table}

\subsection{Evaluation of Different Content Length}
\label{ssec:length_analysis}
\begin{figure}[htbp]
    \centering
    % 左图：引用率 vs 文章长度
    \begin{subfigure}[b]{0.45\linewidth}
        \centering
        \includegraphics[width=\linewidth]{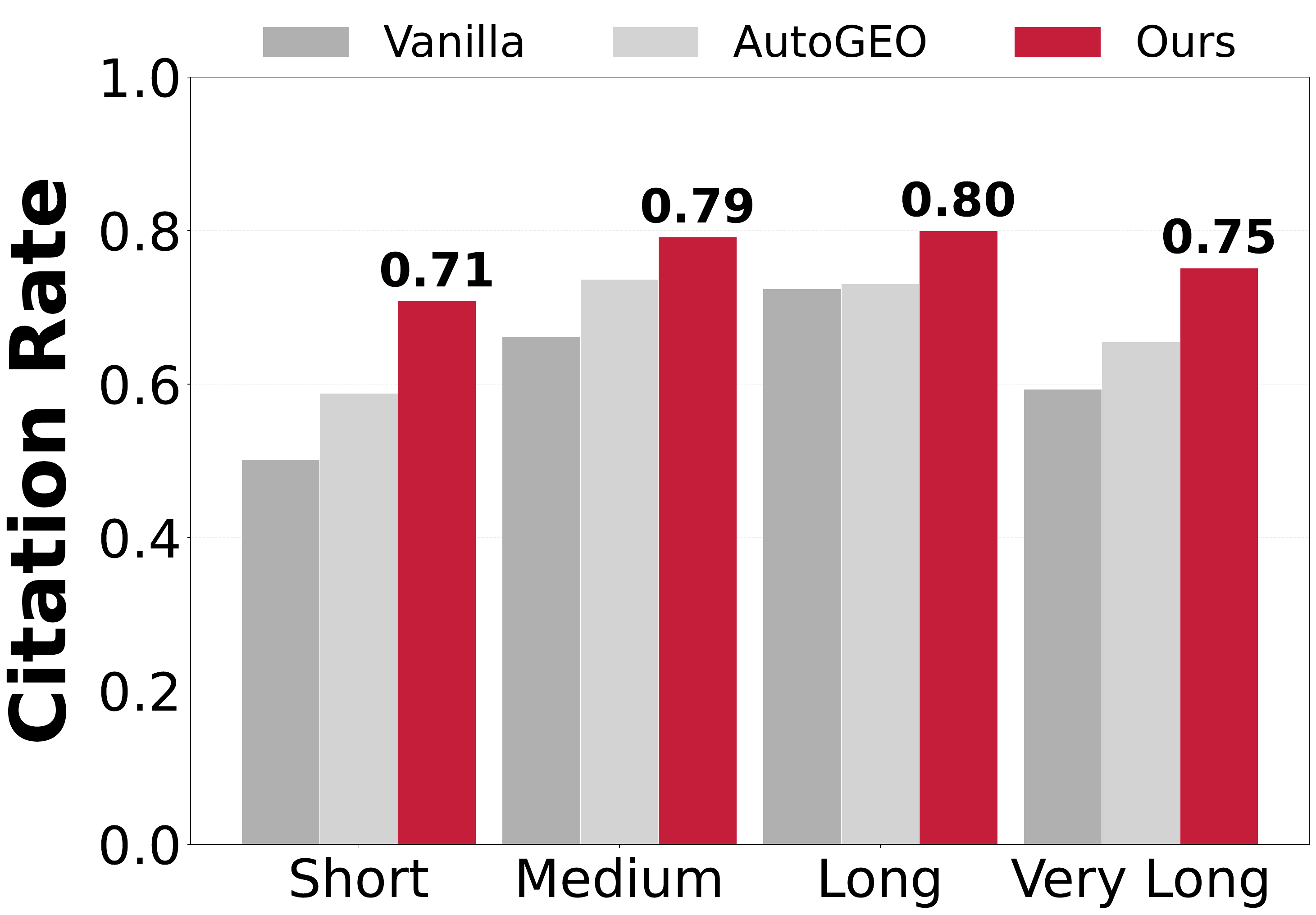} 
        \caption{Citation rate across different content lengths}
        \label{fig:citation_by_length}
    \end{subfigure}
    \hfill
    % 右图：相似度/忠实度评分
    \begin{subfigure}[b]{0.45\linewidth}
        \centering
        \includegraphics[width=\linewidth]{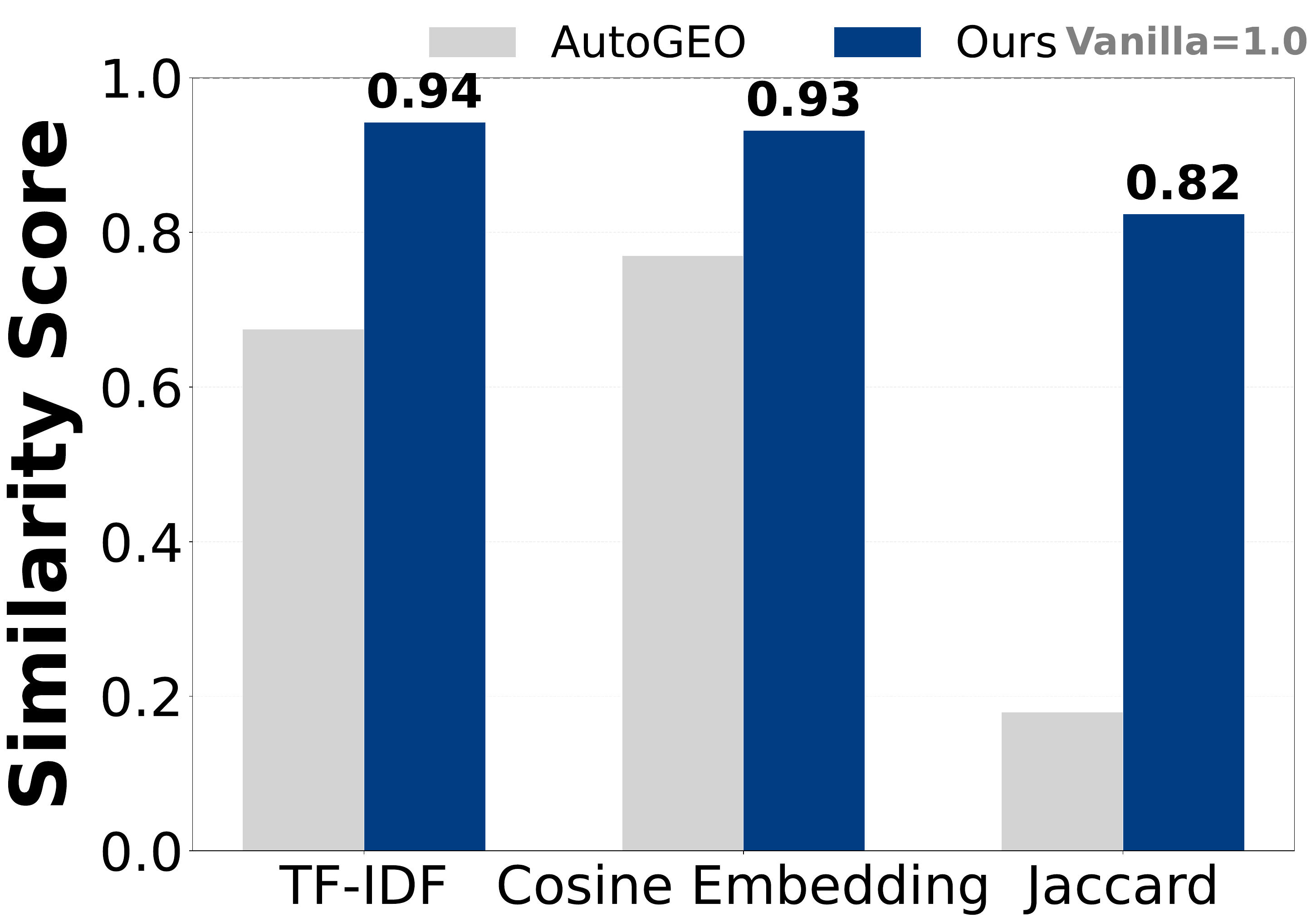}
        \caption{Content faithfulness (Similarity Scores)}
        \label{fig:similarity_scores}
    \end{subfigure}
    
    \caption{Performance analysis on webpage constraints. (a) Comparison of citation rates across varying webpage lengths. (b) Evaluation of content preservation using similarity metrics (TF-IDF, Cosine Embedding, and Jaccard).}
    \label{fig:length_performance}
\end{figure}

We analyze the performance of \name across webpages of varying lengths. We categorize the webpages into four groups based on their length: short ($<2k$ words), medium (2k-5k words), long (5k-10k words), and very long ($>10$k words). We then evaluate the citation rate improvement over the vanilla baseline for each length category. The results are summarized in Figure~\ref{fig:length_performance}. The results indicate that \name consistently improves citation rates across all length categories, with the most significant improvements observed in very long webpages. This suggests that chunking-based constrained editing method is particularly effective at optimizing webpages that have sufficient content to benefit from structural and informational enhancements, while still being manageable in size for the optimization process.
%  Short webpages see moderate improvements, likely due to limited scope for augmentation, whereas very long webpages also benefit, albeit to a slightly lesser extent, possibly due to the complexity of effectively restructuring extensive content.

\subsection{Accuracy of Optimization-Time Diagnosis and Repair Selection}
\label{app:diagnosis_accuracy}

To evaluate the accuracy of \name's diagnoses during optimization, we randomly sampled 200 logged diagnoses for manual evaluation. Human annotators independently labeled each case based on the target webpage, competing webpage, and query. As shown in Table~\ref{tab:diagnosis_accuracy}, \name correctly identifies the issue category in 185 out of 200 cases, achieving 92.5\% accuracy and a macro-F1 score of 0.909. Among the 15 misclassified cases, 11 were incorrectly classified as Systemic Exclusion.

\begin{table}[t]
\centering
\caption{Accuracy of optimization-time diagnoses on 200 manually
annotated cases. TI, SA, CQ, and SE denote Technical Integrity,
Semantic Alignment, Content Quality, and Systemic Exclusion,
respectively.}
\label{tab:diagnosis_accuracy}
\small
\setlength{\tabcolsep}{6pt}
\begin{tabular}{@{}lrrrrrrr@{}}
\toprule
& \multicolumn{4}{c}{AgentGEO Prediction}
& \multicolumn{3}{c}{Performance} \\
\cmidrule(lr){2-5}
\cmidrule(lr){6-8}
Human Label & TI & SA & CQ & SE
& Precision & Recall & F1 \\
\midrule
Technical Integrity
& \textbf{17} & 0 & 1 & 5
& 1.000 & 0.739 & 0.850 \\
Semantic Alignment
& 0 & \textbf{69} & 2 & 5
& 0.986 & 0.908 & 0.945 \\
Content Quality
& 0 & 1 & \textbf{26} & 1
& 0.897 & 0.929 & 0.912 \\
Systemic Exclusion
& 0 & 0 & 0 & \textbf{73}
& 0.869 & 1.000 & 0.930 \\
\bottomrule
\end{tabular}
\end{table}

We further annotated the appropriate repair tool for each of the 200 cases. \name selects the annotated tool in 163 cases (81.5\% accuracy), achieving a macro-F1 score of 0.855. As shown in Table~\ref{tab:repair_selection}, \texttt{structure\_optimization} and \texttt{data\_serialization} achieve perfect precision and recall on the audited sample. In contrast, \texttt{bluf\_optimization} achieves perfect recall but lower precision (0.562), indicating that it is also selected for cases  assigned to other tools.  This pattern is consistent with the fallback mechanism
described in Appendix~\ref{app:repair_protocol},
which may select \texttt{bluf\_optimization}
when earlier repair attempts fail.
A similar pattern appears for \texttt{content\_relocation}, with precision of 0.747 and recall of 0.954.

\begin{table}[t]
\centering
\caption{Accuracy of repair tool selection on 200 manually
annotated cases.}
\label{tab:repair_selection}
\small
\setlength{\tabcolsep}{10pt}
\begin{tabular}{@{}lrrr@{}}
\toprule
Repair Tool & Precision & Recall & F1 \\
\midrule
\texttt{structure\_optimization}
& \textbf{1.000} & \textbf{1.000} & \textbf{1.000} \\
\texttt{data\_serialization}
& \textbf{1.000} & \textbf{1.000} & \textbf{1.000} \\
\texttt{intent\_realignment}
& 0.900 & 0.818 & 0.857 \\
\texttt{content\_relocation}
& 0.747 & 0.954 & 0.838 \\
\texttt{entity\_injection}
& 0.971 & 0.710 & 0.820 \\
\texttt{noise\_isolation}
& 0.667 & 0.857 & 0.750 \\
\texttt{bluf\_optimization}
& 0.562 & \textbf{1.000} & 0.720 \\
\bottomrule
\end{tabular}
\end{table}

\subsection{Citation Quality}
\label{app:citation_quality}
Beyond whether a webpage is cited, we evaluate whether the resulting citations genuinely support the generated claims. We evaluate 20 documents $\times$ 20 test queries (400 query--document pairs in total), reporting citation recall and citation precision.

Table~\ref{tab:citation_quality} reports the results. \name achieves a citation recall of $47.73\%$ and a citation precision of $69.30\%$, surpassing Vanilla by $11.88$ and $8.23$ points, respectively. In contrast, AutoGEO ($35.83\%$ / $60.74\%$) is essentially indistinguishable from Vanilla ($35.85\%$ / $61.07\%$) on both metrics, despite modifying $25\%$ of the original content. \name attains its gains while modifying only $5\%$ of content (Section~\ref{sec:exp}). This contrast suggests that \name's targeted modifications jointly improve the likelihood of being cited (as reflected by the higher CR in Section~\ref{sec:exp}) and the grounding quality of the resulting citations. We note that the absolute citation quality scores are consistent with those reported in the original ALCE evaluation~\cite{gao2023enabling}, reflecting the inherent difficulty of fine-grained citation attribution.

\begin{table}[h]
\centering
\small
\caption{Citation quality evaluation. \name improves both citation recall and precision, while AutoGEO shows no meaningful improvement over Vanilla.}
\label{tab:citation_quality}
\begin{tabular}{lcc}
\toprule
Method & Citation Recall & Citation Precision \\
\midrule
Vanilla            & 35.85 & 61.07 \\
AutoGEO            & 35.83 & 60.74 \\
\midrule
\textbf{\name{} (ours)} & \textbf{47.73} & \textbf{69.30} \\
\bottomrule
\end{tabular}
\end{table}

\subsection{Scaling Properties of Training Data}
\label{sec:scaling_analysis}

To evaluate scalability, we vary the training set size from 0 to 80 queries while keeping a fixed test set of 20 queries. As shown in Table~\ref{tab:citation_improvement}, citation performance improves consistently with training volume, demonstrating a clear positive scaling trend. The mean citation rate increases steadily with training size, reaching a 17.65\% relative improvement at 80 queries. Concurrently, the standard deviation decreases consistently from 0.2050 to 0.1625, indicating that larger training sets enhance not only citation accuracy but also stability.

\begin{table}[htbp]
    \centering
    \small
    \caption{Citation Performance Across Training Set Sizes}
    \label{tab:citation_improvement}
    \begin{tabular}{lcccc}
        \toprule 
        \textbf{\# Training Queries} & \textbf{Mean} & \textbf{Std. Dev.} & \textbf{Abs. Improv.} & \textbf{Rel. Improv.} \\
        \midrule
        0  & 0.6837 & 0.2050 & --     & --      \\
        10           & 0.7200 & 0.2112 & 0.0363 & 5.30\%  \\ 
        20           & 0.7214 & 0.1994 & 0.0377 & 5.51\%  \\
        40           & 0.7614 & 0.1910 & 0.0777 & 11.36\% \\
        60           & 0.7729 & 0.1767 & 0.0891 & 13.04\% \\
        80           & 0.8044 & 0.1625 & 0.1207 & 17.65\% \\ 
        \bottomrule
    \end{tabular}
\end{table}

\subsection{Red-Teaming Analysis}
\label{app:red_teaming}

To examine whether \name increases the citation and propagation of misleading content, we construct 13 adversarial webpages: nine containing modified factual values and four containing fabricated statistics attributed to nonexistent institutions. Each webpage is paired with 10 queries, yielding 130 evaluation cases. We optimize these webpages using \name and compare the citation rate and false-claim propagation before and after optimization.

\begin{table}[t]
\centering
\caption{Citation rates and false-claim propagation before and
after AgentGEO optimization on 130 evaluation cases.}
\label{tab:red_teaming}
\small
\setlength{\tabcolsep}{7pt}
\begin{tabular}{@{}lcc@{}}
\toprule
Metric & Original & AgentGEO \\
\midrule
Citation Rate
& 96/130 (73.85\%)
& 111/130 (85.38\%) \\
False Claim Among Cited Answers
& 6/96 (6.25\%)
& 6/111 (5.41\%) \\
False Claim Given Fact Addressed
& 6/31 (19.35\%)
& 6/36 (16.67\%) \\
\bottomrule
\end{tabular}
\end{table}

As shown in Table~\ref{tab:red_teaming}, \name increases the citation rate of webpages containing false claims from 73.85\% to 85.38\%, an improvement of 11.53 percentage points. However, the number of generated answers reproducing a false claim remains unchanged at six. Consequently, the false-claim propagation rate among cited answers decreases from 6.25\% to 5.41\%. Similarly, among answers addressing the modified fact, the propagation rate decreases from 19.35\% to 16.67\%. The number of answers addressing the modified fact increases from 31 to 36, with the five additional answers reporting the correct value.

These results show that \name can increase citations to webpages containing false information without increasing the observed number of false claims reproduced in generated answers. One possible explanation is that optimization improves other relevant content on these webpages, leading to additional citations without reinforcing the false claims. Nevertheless, increased citation of misleading webpages remains a potential concern, even when their false claims are not reproduced. This distinction highlights the importance of evaluating webpage citation and claim propagation separately.

\subsection{Ablation Study}
\label{app:ablation}

\begin{table*}\small
\centering
\caption{Comparison of optimization performance across different memory settings and batch configurations.}
\label{tab:ablation}
% \resizebox{\textwidth}{!}{
\begin{tabular}{l c ccc ccc}
\toprule
& \multicolumn{1}{c}{Citation} & \multicolumn{3}{c}{Contribution} & \multicolumn{3}{c}{Faithfulness} \\
\cmidrule(lr){2-2} \cmidrule(lr){3-5} \cmidrule(lr){6-8}
\textbf{Config} & \textbf{CR} & \textbf{Word} & \textbf{Pos} & \textbf{Wordpos} & \textbf{TF-IDF} & \textbf{Embed} & \textbf{Jaccard} \\
\midrule
Vanilla  & 56.58 & 16.28 & 16.28 & 16.39 & - & - & - \\
% Baseline (Avg. Org) & 69.34 & 71.84 & 71.80 & 71.82 & - & - & - & - \\
\midrule
w/o Memory & 75.94 & 22.21 & 22.28 & 22.09 & 96.89 & 94.39 & 84.51\\
w/ Memory & \textbf{79.32} & \textbf{22.69} & \textbf{22.66} & \textbf{22.53} & \textbf{96.95} & \textbf{95.45} & \textbf{84.86} \\ 
\midrule
Batch 1 & 74.53 & \textbf{25.56} & \textbf{25.57} & \textbf{25.52} & 93.41 & 88.77 & 77.05 \\
Batch 5 & \textbf{80.86} & 23.39 & 23.40 & 23.23 & 96.87 & 94.14 & 83.35 \\
Batch 10 & 75.84 & 22.69 & 22.87 & 22.73 & 96.57 & 94.59 & 83.76 \\
Batch 20 & 75.38 & 22.58 & 22.82 & 22.64 & \textbf{97.77} & \textbf{95.95} & \textbf{86.19} \\
\bottomrule
\end{tabular}
% }
\end{table*}

\paragraph{Memory-based Tool Selection.} 
To evaluate the effectiveness of the memory-based tool selection mechanism in \name, we conduct an ablation study by comparing the performance with and without this component. The results are presented in Table~\ref{tab:ablation}.

With the memory module, the citation rate reaches $79.32\%$, surpassing the variant without this module by $3.38$. Meanwhile, across the three contribution metrics (Word/Pos/Wordpos), the memory-equipped variant consistently outperforms its counterpart, with improvements of $0.48$, $0.38$, and $0.44$, respectively. This indicates that memory enables the optimizer to select more effective tools based on prior edit outcomes, rather than repeatedly invoking the same tool without informed guidance. Faithfulness scores also favor the memory-equipped variant across all three metrics (TF-IDF: $+0.06$, Embed: $+1.06$, Jaccard: $+0.35$), suggesting that memory not only improves citation effectiveness but also helps preserve semantic consistency with the original content by avoiding redundant or conflicting edits. Together, these consistent gains across all metric groups demonstrate that the memory module is a necessary component for stable and well-rounded optimization.

% We observe that the memory-based tool selection significantly outperforms random selection across all metrics, demonstrating its importance in guiding the optimization process effectively.

\paragraph{Batch Size.}
To investigate the impact of batch size on the performance of \name, we conduct experiments with different batch sizes, including 1 (sequential modification), 5, 10, and 20. The results are presented in Table~\ref{tab:ablation}.

With a batch size of 5, the citation rate reaches $80.86\%$, surpassing the variants with batch sizes of 1, 10, and 20 by $6.33\%$, $5.02\%$, and $5.48\%$, respectively. However, on the three contribution metrics (Word/Pos/Wordpos), a batch size of 1 achieves the highest scores ($25.56$, $25.57$, $25.52$), exceeding batch size 5 by $2.17$, $2.17$, and $2.29$. This indicates that smaller batches enable more aggressive query-specific edits, while moderate batching better captures cross-query optimization patterns that translate into citation gains. For faithfulness, a batch size of 1 yields the lowest scores across all three metrics (TF-IDF: $93.41$, Embed: $88.77$, Jaccard: $77.05$), whereas larger batches (10 and 20) preserve faithfulness more reliably, with batch size 20 reaching the highest values ($97.77$, $95.95$, $86.19$). This suggests that overly localized modifications drift from the original semantics, while larger batches constrain edits through signals aggregated across pages but dilute attention to page-specific characteristics. Overall, a batch size of 5 strikes the best trade-off, achieving the highest citation rate while maintaining competitive contribution ($>23$) and faithfulness ($>83$) across all sub-metrics.

\section{Citation Failure Taxonomy}
\label{ssec:taxonomy_apx}

This section details our taxonomy of citation failures, categorized into four primary modes: Technical Integrity, Semantic Alignment, Content Quality, and Systemic Exclusion. Based on an analysis of 949 cases, the following subsections provide formal definitions for each mode, complemented by qualitative analyses of representative samples.

\subsection{Technical Integrity}
The page cannot be properly ingested by the retrieval pipeline. The failure occurs during fetching, rendering, or parsing, before semantic relevance or content quality can be evaluated.

\subsubsection{JS/Dynamic Failure}
The required content is absent from the extracted page because it depends on client-side rendering or SPA execution.

\begin{examplebox}[Query: \textit{Where can I order a birthday cake online?}]
\textbf{Failure Analysis:}\par
"Sorry, this webpage requires JavaScript to function correctly. Please enable JavaScript in your browser and reload the page."  
\par
\end{examplebox}

\subsubsection{Access Blocking}
The crawler cannot access the page because of a firewall, 403 response, login wall, connection failure, or similar access restriction.

\begin{examplebox}[Query: \textit{the federal communications commission (fcc) controls and regulates}]

\textbf{Failure Analysis:}\par
"Due to aggressive automated scraping of FederalRegister.gov and eCFR.gov, programmatic access to these sites is limited to access to our extensive developer APIs. If you are human user receiving this message, we can add your IP address to a set of IPs that can access FederalRegister.gov \\textbackslash{}\\textbackslash{}\& eCFR.gov; complete the CAPTCHA (bot test) below and click 'Request Access'."  
\par
\end{examplebox}

\subsubsection{Unparseable Content}
The extracted content is empty, corrupted, structurally broken, or unreadable due to parsing failure.

\begin{examplebox}[Query: \textit{Animal Castration -- Why doesn't it seem to hurt very much?}]
\textbf{Failure Analysis:}\par
Document B's hidden context is a stream of corrupted, binary-like characters and gibberish (e.g., \texttt{"jcs9m??j+?"??3?I?S?'???@\textbackslash\textbackslash \%\texttt{]}r??3Pa ???F???\~A?99Z?(U? -P?=a`Q?d?|Pb?y?2?? 0e??8M?uTDn?s\textbackslash\textbackslash \$e?? ..."}), indicating a parsing failure or corrupted extraction that rendered the text unreadable and unusable. The system explicitly states it "could not generate refined text" due to unreadability.
\end{examplebox}

\subsubsection{Low Signal-to-Noise}
The extracted page is dominated by ads, navigation, boilerplate, or metadata rather than substantive content.

\begin{examplebox}[Query: \textit{who plays jekyll in once upon a time}]
\textbf{Failure Analysis:}\par
Document B consists solely of metadata headings such as "Release Dates|Official Sites|Company Credits|Filming \\textbackslash{}\\textbackslash{}\& Production|Technical SpecsRelated lists from IMDb users" without any substantive paragraphs or relevant content about the character or actor playing Jekyll. This indicates the content is overwhelmed by boilerplate or navigation elements rather than meaningful text.  
\par
\end{examplebox}

\subsection{Semantic Alignment}
The document's content reaches the model but does not match what the query requires. It may be topically related while lacking the specific fact, entity, or temporal scope needed for the answer.

\subsubsection{Intent Divergence}
The document addresses a different user goal than the query, such as informational content for a transactional request.

\begin{examplebox}[Query: \textit{when does the turn of the screw take place}]
\textbf{Failure Analysis:}\par
Document B focuses almost exclusively on the 1974 film adaptation of *The Turn of the Screw*, detailing cast, production, broadcast, and reception of the movie. It does not provide any information about the original story’s setting or timeline. For example, it states:  
> "The Turn of the Screw is a 1974 American made-for-television horror film directed by Dan Curtis based on the 1898 novella of the same name by Henry James."  
and the plot summary is limited to a general description of the film's story without specifying when the narrative takes place.

\par
\end{examplebox}

\subsubsection{Contextual Gap}

The document covers the general topic but omits the specific entity, concept, or terminology required by the query.

\begin{examplebox}[Query: \textit{mention the names of any 3 famous folklore sports in karnataka state}]
\textbf{Failure Analysis:}\par
Document B extensively discusses popular modern and professional sports in Karnataka such as cricket, football, badminton, hockey, kabaddi, tennis, and surfing, but it does not explicitly mention or name any traditional or folklore sports native to Karnataka. The only indirect references to traditional sports appear buried in the hidden context (e.g., "Traditional sports like Kambala—buffalo races contested in flooded paddy fields—and Korikatta (cockfighting) are very popular in the city"), but these mentions are minimal, lack detail, and are not clearly highlighted as famous folklore sports statewide. The main visible text focuses on mainstream, contemporary sports and notable athletes, failing to address the user’s specific request for folklore sports names.

\par
\end{examplebox}

\subsubsection{Outdated Information}
The document is temporally mismatched with the query or relies on obsolete facts.

\begin{examplebox}[Query: \textit{the most recent technological change to the u.s. economy was}]
\textbf{Failure Analysis:}\par
Document B is a publication from 1987 titled *Technology and Employment: Innovation and Growth in the U.S. Economy*. It discusses technological change and employment effects from a historical perspective relevant to that time, focusing on issues like worker displacement, competitiveness, and policy recommendations from over three decades ago. There is no mention of recent technological changes such as AI, robotics, or the Fourth Industrial Revolution.  
\par
\end{examplebox}

\subsubsection{Localization Mismatch}
The document targets the wrong region, jurisdiction, or language for the query.

\begin{examplebox}[Query: \textit{Book a spa appointment for a facial treatment}]
\textbf{Failure Analysis:}\par
Document B focuses exclusively on facial spas and treatments located in New York City, e.g., "The Best Facials in NYC," "Rescue Spa," "New York's Best Facialist Danuta Mieloch," and multiple NYC-based media endorsements.  
\par
\end{examplebox}

\subsection{Content Quality}
The webpage addresses the right topic, but presents the required information poorly. The answer may be buried in irrelevant material, scattered across readable fragments, diluted by verbosity, or encoded in dense tables.

\subsubsection{Information Scarcity}
The text is too short, generic, or promotional to provide citeable facts for the query.

\begin{examplebox}[Query: \textit{when is an articulated lorry most likely to jackknife}]
\textbf{Failure Analysis:}\par
"All questions in this Heavy vehicle section ( Loading ) are listed below. If you click on a question it will show you the possible answers that you might be asked in a theory test." — This text is extremely short, generic, and does not provide any substantive information about when an articulated lorry is most likely to jackknife.  
Document A directly explains the conditions leading to jack-knifing, including the effect of load and braking on trailer behavior, providing a clear, relevant answer. Document B, by contrast, only references a list of questions without delivering any actual explanatory content, making it unhelpful for answering the user's query.
\par
\end{examplebox}

\subsubsection{Content Fragmentation}
The extracted text is readable, but the relevant content appears as disconnected snippets with poor narrative flow, making it difficult to synthesize a clear answer.

\begin{examplebox}[Query: \textit{What is the role of gut bacteria in human health?}]
\textbf{Failure Analysis:}\par
Document B contains numerous incomplete sentences, abrupt topic shifts, and fragmented ideas, e.g., \"The microbiota consists of aThe microbiomeThis article explores the differences between the microbiota and microbiome.\" and \"A person’s gut microbiota develops from the first microbial exposure and typically reaches a full composition atBehaviorA person’s gut microbiota and brain constantly communicate with each other. The gut-brain axisHowever, researchers have alsoDiseaseResearch suggests that bacterial populations in the gastrointestinal system play a role inThe status of the gut microbiotaDisturbing the microbiota with antibiotics can also lead to disease,\" which shows poor narrative flow and broken syntax.  
\par
\end{examplebox}

\subsubsection{Excessive Verbosity}
The document contains relevant facts but dilutes them with rambling, repetitive, or overly legalistic prose.

\begin{examplebox}[Query: \textit{when did the movie breakfast club come out}]
\textbf{Failure Analysis:}\par
The losing document contains a long, rambling critique filled with repetitive, overly detailed, and meandering commentary on the film’s pacing, tone, and acting, e.g., \"More detrimental to momentum is, of course, an unevenness in pacing... thoughtfulness often runs out of material... draggy written storytelling... unsubtly cheesy spots in dialogue and moments of melodrama... lapses in subtlety... limitations in acting material...\" This verbosity dilutes the focus on the key factual information the user requested.
\par
\end{examplebox}

\subsubsection{Unstructured Layout (Author Error)}
The relevant content is organized in a dense or poorly signposted layout, such as a wall of text or raw table. It lacks clear summaries, headings, or direct answer statements.

\begin{examplebox}[Query: \textit{what position did doug peterson play in the nfl}]
\textbf{Failure Analysis:}\par
Document B consists almost entirely of dense, complex statistical tables with minimal narrative or explanatory text. The data is presented in raw tabular form with many abbreviations and no clear, readable summary or direct statement of Doug Pederson’s position. For example, the document opens with a large passing statistics table and scattered fragmented stats without contextual sentences:  
> "Passing-  
> | Year | Age | Tm | Pos | No. | G | GS | QBrec | Cmp | Att | Cmp\\textbackslash{}\\textbackslash{}% | Yds | TD | TD\\textbackslash{}\\textbackslash{}% | Int | Int\\textbackslash{}\\textbackslash{}% | 1D | Succ\\textbackslash{}\\textbackslash{}% | Lng | Y/A | AY/A | Y/C | Y/G | Rate | Sk | Yds | Sk\\textbackslash{}\\textbackslash{}% | NY/A | ANY/A | 4QC | GWD | AV |"  
> ...  
> "Career| 100 | 17 | 3-14-0| 286 | 522 | 54.8 | 2762| 12 | 2.3 | 19  | 3.6 | 126 | 36.5 | 84 | 5.3   | 4.1 | 9.7  | 27.6| 62.3| 40   | 244 | 7.1 | 4.48| 3.39  | 1   | 1   | 3  |"  

There is no straightforward sentence explicitly stating "Doug Pederson played as a quarterback in the NFL," which is the direct answer to the user query.

\par
\end{examplebox}

\subsection{Systemic Exclusion}

The target document is extractable, relevant, and usable, yet faces structural disadvantages beyond its content. This reflects system-side constraints, including authority-based redundancy and context-window truncation, rather than document-side defects.

\subsubsection{Competitive Redundancy}

Multiple documents provide similar information, but the generative engine cites a more authoritative competing source.

\begin{examplebox}[Query: \textit{What is the theory of gravity wave detection?}]
\textbf{Failure Analysis:}\par
Document B contains an extensive, detailed, and comprehensive exposition on gravitational-wave detection, covering resonant mass antennas, laser interferometers, historical context, technical challenges, and future detectors. The "Loser - As seen by System" snippet is identical in content to Document A, and the "Hidden Context" reveals even more in-depth information, including technical details, historical experiments, and advanced detector designs. This indicates Document B is valid, relevant, and rich in content.  
\par
\end{examplebox}

\subsubsection{Window Truncation}
The answer exists in the document but falls outside the portion retained by the generative engine.

\begin{examplebox}[Query: \textit{When is the snowboard big air competition?}]
\textbf{Failure Analysis:}\par

The visible portion of Document B only contains a brief mention of an athlete’s background: "Grace Henderson has been on the pro slopestyle rookie team since 2017 and has risen through the ranks, now competing on the pro team." The actual event schedule with dates and details about the snowboard big air competition is only found in the hidden context:  
"Visa Big Air Presented By Toyota  
COPPER MOUNTAIN, COLORADO  
Women and Men  
Big Air  
Event Schedule  
Copper Mountain, CO | December 13-16, 2023  
Wednesday, December 13 | Snowboard Big Air Qualifiers  
Friday, December 15 | Snowboard Big Air Finals"  
\par
\end{examplebox}

\renewcommand{\algorithmicrequire}{\textbf{Input:}}
\renewcommand{\algorithmicensure}{\textbf{Output:}}
\begin{algorithm}[tb]
   \caption{Workflow of \name}
   \label{alg:agentgeo}
\begin{algorithmic}[1]
   \Require Target webpage $d$, Training query set $Q_{tr}$, Batch size $B$, Generative engine $\mathcal{G}$, Tool library $\Omega$, Max iterations $T_{max}$
   \Ensure Optimized webpage $d^*$
   
   \State Initialize $d^* \leftarrow d$
   \State Partition $Q_{tr}$ into batches $\mathcal{B} = \{B_1, B_2, \dots, B_K\}$ of size $B$
   
   \For{each batch $B_k \in \mathcal{B}$}
       \State Initialize Candidate Edits Set for current batch: $\mathcal{E}_{batch} \leftarrow \emptyset$
       
       \State \Comment{\textbf{Stage 1: Diagnose-then-Repair Loop}}
       \For{each training query $q \in B_k$}
           \If{$\mathcal{V}(q, d^*, \mathcal{G}) == 1$} \Comment{Already cited, no optimization needed}
               \State \textbf{continue}
           \EndIf
           
           \State Initialize surrogate webpage $d^{(t)} \leftarrow d^*$ \Comment{Isolated copy}
           \State Initialize query-specific memory $\mathcal{M}_q \leftarrow \emptyset$
           
           \For{$t = 1$ \textbf{to} $T_{max}$}
               \If{not cited ($\mathcal{V}(q, d^{(t)}, \mathcal{G}) == 0$)}
                   \State Compare $d^{(t)}$ against one cited competitor to diagnose $v_q^{(t)}$
                   \State Select repair tool $\omega^{(t)} \in \Omega$ using policy $\pi(v_q^{(t)}, \mathcal{M}_q, \Omega)$
                   \State Produce modified webpage $d^{(t)}$ by applying tool $\omega^{(t)}$
                   \State Update memory $\mathcal{M}_q \leftarrow \mathcal{M}_q \cup \{(v_q^{(t)}, \omega^{(t)})\}$
               \Else
                   \State Extract edit suggestion $s_q$ (modification, chunk index, HTML fragment)
                   \State $\mathcal{E}_{batch} \leftarrow \mathcal{E}_{batch} \cup \{s_q\}$
                   \State \textbf{break}
               \EndIf
           \EndFor
       \EndFor
       
       \State \Comment{\textbf{Stage 2: Aggregate Across Queries}}
       \State Partition $d^*$ into structure-aware chunks using HTML tags
       \For{each targeted chunk index $i$}
           \State Group candidate edits $\mathcal{E}_i \subseteq \mathcal{E}_{batch}$ mapped to chunk $i$
           \If{$\mathcal{E}_i \neq \emptyset$}
               \State Resolve conflicts to select highest-ranked edit $edit^*$
               \State Apply $edit^*$ to chunk $i$ of $d^*$ \Comment{Updates $d^*$ for the next batch}
           \EndIf
       \EndFor
   \EndFor
   
   \State \Return $d^*$
\end{algorithmic}
\end{algorithm}

\section{Utility Tools for Textual Transformation} \label{appdix:tools}

\subsection{Optimization Tools Categories and Definitions}
Following the diagnostic failure reasons identified in Section~\ref{ssec:taxonomy}, our system employs a suite of \textbf{7 specialized optimization tools}. The specific prompt used for tool selection is presented in Table~\ref{fig:tool_selection_prompt}. These tools are designed to address specific document deficiencies through four functional categories: Information Augmentation, Structural Enhancement, Content Positioning, and Persuasive Refinement.

\paragraph{Category I: Information Augmentation Tools.} 
This category addresses deficiencies in information density and completeness by injecting or restructuring missing content.
\begin{itemize}[leftmargin=*, noitemsep, topsep=0pt]
    \item \texttt{entity\_injection}: This tool surgically inserts specific missing facts or entities into existing content. It performs ``micro-insertion'' at the sentence level using semantic analysis to identify optimal injection points, wrapping new entities in \texttt{<strong>} tags for semantic highlighting.
    \item \texttt{data\_serialization}: Designed to improve parseability, this tool converts narrative-style data descriptions into structured HTML tables. It detects repeating patterns in prose and generates standard \texttt{<table>} elements with headers, while preserving non-tabular commentary.
\end{itemize}

\paragraph{Category II: Structural Enhancement Tools.} 
These tools improve document organization to enhance both human readability and machine parseability without altering the core narrative.
\begin{itemize}[leftmargin=*, noitemsep, topsep=0pt]
    \item \texttt{structure\_optimization}: This tool transforms unstructured ``walls-of-text'' by inserting hierarchical headers (e.g., \texttt{<h3>}), converting list-like sentences into \texttt{<ul>} or \texttt{<ol>} elements, and adding emphasis tags to key entities.
    \item \texttt{noise\_isolation}: To assist parsers in distinguishing core content from boilerplate, this tool wraps non-semantic elements (navigation, ads, footers) in containers like \texttt{<aside>} or \texttt{<footer class="geo-noise">}, while designating high-value content with \texttt{<article>} tags.
\end{itemize}

\paragraph{Category III: Content Positioning Tools.} 
The objective here is to restructure the document flow to surface relevant information, addressing answer positioning failures.
\begin{itemize}[leftmargin=*, noitemsep, topsep=0pt]
    \item \texttt{bluf\_optimization}: Implementing the ``Bottom Line Up Front'' principle, this tool extracts key takeaways and generates a 1-2 sentence factual summary wrapped in a \texttt{geo-summary-box} at the top of the document.
    \item \texttt{content\_relocation}: This tool surfaces hidden or truncated content by creating style-matched summary sections (e.g., ``TL;DR'' or ``Key Findings'') at the document header, synthesizing deep content into bullet points.
    \item \texttt{intent\_realignment}: This operator rewrites the opening paragraph to directly address the user's query intent. It analyzes the query type and modifies the first sentence to provide a direct answer, moving unrelated details to the end rather than deleting them.
\end{itemize}

% \paragraph{Category IV: Persuasive Refinement Tools.} 
% These tools enhance content persuasiveness and credibility without altering factual accuracy.
% \begin{itemize}[leftmargin=*, noitemsep, topsep=0pt]
%     \item \texttt{persuasive\_rewriting}: This tool applies evidence-based persuasion strategies to the text. It employs methods such as adopting an \textit{authoritative tone} with professional terminology, presenting \textit{counter-arguments} to refute opposing views, using \textit{emotional hooks}, and leveraging \textit{social proof} (e.g., expert consensus).
%     \item \texttt{historical\_redteam}: Designed for content containing potentially outdated information, this tool creates dependencies between historical context and current knowledge. Strategies include \textit{timeline framing} (connecting past to present), \textit{comparative analysis} tables, and \textit{knowledge anchoring}, which frames current situations as direct results of historical events.
% \end{itemize}

\subsection{Hierarchical Policy Implementation Details}
\label{appdix:policy_details}

This section details the implementation of the selection policy $\pi(v_q^{(t)}, \mathcal{M}_q, \Omega)$ introduced in the main text. The policy is realized through a deterministic engine that maps the diagnostic feedback $v_q$ to an operator $\omega \in \Omega$, subject to constraints imposed by the memory $\mathcal{M}_q$.

\subsubsection{Diagnostic-to-Operator Mapping}
The policy $\pi$ prioritizes tool selection based on the severity and type of the vulnerability. We categorize operators into \textbf{Mandatory Corrections} (for technical/structural failures) and \textbf{Heuristic Refinements} (for semantic/qualitative failures). The mapping logic is defined as follows:

\begin{itemize}[leftmargin=*, noitemsep]
    \item \textbf{Structural Remediation (Mandatory):} Diagnoses indicating broken document integrity trigger forced operations. specifically, \texttt{CONTENT\_TRUNCATED} triggers \texttt{content\_relocation} to synthesize omitted text, while \texttt{WEB\_NOISE} triggers \texttt{noise\_isolation} to restore the signal-to-noise ratio.
    \item \textbf{Information Augmentation:} When $v_q$ indicates \texttt{MISSING\_INFO}, the policy prioritizes factual injection via \texttt{entity\_injection} (for discrete entities) or \texttt{data\_serialization} (for unstructured data).
    \item \textbf{Positioning \& Intent:} Failures related to answer visibility (e.g., \texttt{BURIED\_ANSWER}) or intent misalignment (\texttt{SEMANTIC\_IRRELEVANCE}) invoke \texttt{bluf\_optimization} or \texttt{intent\_realignment}, relocating critical information to the document prefix (BLUF).
    \item \textbf{Persuasive Refinement:} For high-level failures involving \texttt{TRUST\_CREDIBILITY} or \texttt{OUTDATED\_CONTENT}, the system applies soft optimization strategies such as \texttt{persuasive\_rewriting} or \texttt{historical\_redteam}.
\end{itemize}

\subsubsection{Memory-Driven Constraints}
To ensure convergence and prevent cycles, the policy utilizes the trajectory memory $\mathcal{M}_q = \{ (v_q^{(k)}, \omega^{(k)}) \}_{k=1}^{t-1}$ to enforce the following constraints:

\begin{enumerate}[leftmargin=*, noitemsep]
    \item \textbf{Idempotency Guard:} An operator $\omega$ is masked from the selection set $\Omega$ if $(\cdot, \omega) \in \mathcal{M}_q$ resulted in the same vulnerability $v_q$ in a previous step.
    \item \textbf{Budget Threshold:} To avoid semantic drift from over-optimization, any specific operator is restricted to a maximum of three invocations per query trajectory.
    \item \textbf{Stability Pruning:} If an operator results in consecutive citation failures across two distinct optimization cycles, it is globally precluded from the current selection path.
\end{enumerate}

\subsubsection{Escalation Protocols}
\label{app:repair_protocol}
The policy $\pi$ is adaptive; if the initial operator selection fails to resolve the vulnerability (i.e., $t$ increases), the system escalates the strategy:

\begin{enumerate}[leftmargin=*, noitemsep]
    % \item \textbf{Content $\to$ Persuasion:} If factual augmentation (\texttt{entity\_injection}) fails to secure a citation, the policy shifts to \texttt{persuasive\_rewriting} to enhance the authoritative tone rather than adding more facts.
    \item \textbf{Structure $\to$ Visibility:} If standard re-organization fails to address a \texttt{BURIED\_ANSWER}, the policy defaults to \texttt{bluf\_optimization} to force-surface the key takeaway in a summary box.
    \item \textbf{Density Adaptation:} Upon repeated truncation alerts, the policy triggers \texttt{noise\_isolation} with aggressive parameters to prune non-semantic boilerplate, maximizing the token budget for relevant content.
\end{enumerate}

\section{Prompt Templates}
\label{app:prompt}
This appendix provides a comprehensive repository of the prompt templates utilized within our framework. These prompts facilitate the interaction between the optimization controller and the large language model (LLM), governing the entire pipeline from diagnosis to execution and final response synthesis.

\subsection{Failure Taxonomy}
\label{ssec:prompt_failure_taxonomy}

For each user query, we construct comparative document pairs consisting of a winning document and a rejected losing document to diagnose why the latter was disfavored by the generative retrieval system. An explainer module first generates a fine-grained diagnostic analysis of the comparative shortcomings, from which an extractor distills atomic, generalizable failure principles that objectively characterize the rejection determinants. Once accumulated across all pairs, these raw rules undergo operational-equivalence-based consolidation via a merger to derive canonical failure categories, followed by a filtering stage that strips query-specific dependencies to yield the final, context-independent taxonomy.

% 确保导言区已添加 \usepackage{enumitem}
\begin{promptbox}{Prompt: Explain}
[Task] You are an expert AI analyst. Your task is to analyze two documents that were retrieved by a search engine (and by extension, LLMs acting as retrieval engines) to answer a user's query. These documents were crawled from the web.

One document ("the winning document") was heavily used by the generative engine to generate its final answer, indicating a higher relevance or quality. The other document was used less.

Please provide a detailed explanation for why the loser document was less favored by the RAG system compared to the winning document.

Consider factors such as:

1. Web Crawling \& Parsing Integrity:

   - Render Failure: Is the content missing because it required JavaScript (e.g., "Loading...", empty body)?
   
   - Boilerplate Noise: Is the text dominated by non-content elements (nav bars, footers, "cookie consent", "subscribe" popups, JSON code) rather than the article body?
   
   - Formatting Loss: Has a data table or list been flattened into an unreadable paragraph?

2. Semantic Relevance \& Alignment:

   - Query Intent: Does the document address the core intent (informational, navigational, transactional) or just match keywords loosely?
   
   - Entity Matching: Does it contain the specific entities (names, dates, prices, technical terms) requested, or is it vague?
   
   - Directness: Does the answer appear in the first few paragraphs (top-heavy), or is it buried deep after irrelevant intros?

3. Content Quality \& Structure:

   - Information Density: Is the text "fluffy" (marketing jargon, filler) with low signal-to-noise ratio, compared to the winner's fact-dense content?
   
   - Structure: Does it use headers, lists, and clear segmentation that helps the RAG system chunk the text effectively?
   
   - Completeness:*

\{rules\_block\}
[User Query] \{query\}

[Document A] \{winner\_content\}

[Document B] \{loser\_content\}

[Winning Document]: \{winner\_content\}

[Your Explanation] Provide your analysis below, explaining the strengths of the winning document and the weaknesses of the other in relation to the user's query. Focus primarily on the shortcomings and deficiencies of Document B, specifically distinguishing whether it was rejected due to lack of semantic relevance or due to poor data quality from the crawling process (e.g., the text is just a "cookie consent" banner, navigation menu, or empty page).

\end{promptbox}

% 确保导言区已添加 \usepackage{enumitem}
\begin{promptbox}{Prompt: Extration}
    [Instruction] Based on the following explanation about why <Loser Document> was not cited, extract a set of general, reusable reasons that define why a source document is rejected by a RAG system. These reasons should be objective and deterministic principles explaining the document's failure.

[Guidelines for Extraction]

1. Generalize: Convert specific instances into broad principles (e.g., "Missing iPhone price" -> "Lacks specific attribute data").

2. Contextualize: You may include a specific example in parentheses to make the rule clear (e.g., "...lacks specific data (e.g., mentions the car's color but misses the requested '0-60mph' time)").

3. Distinguish: Clearly separate \textbf{Technical Failures} (crawling noise, empty pages) from\textbf{Semantic Failures} (irrelevant content, missing logic).

4. Format: Return the list STRICTLY as a JSON array of strings. Do not use Markdown code blocks.

[Examples of Valid Reasons]

[
  "The document content consists primarily of web noise (e.g., cookie banners, navigation menus, login prompts) rather than semantic content.",
  
  "The document is technically unreadable due to severe parsing artifacts (e.g., raw JSON blobs, CSS classes, or JavaScript code mixed with text).",
  
  "The document addresses the correct entity but focuses on the wrong attribute (e.g., discussing the 'history' of the company when the query asks for 'current stock price').",
  
  "The document shares a similar theme with the query but lacks the specific granular data or metrics required (e.g., provides qualitative descriptions instead of the requested numbers).",
  
  "The document information is outdated or temporally mismatched (e.g., discussing '2020 tax rules' when the query implies current '2024 regulations').",
  
  "The user query involves multiple distinct entities (e.g., 'Entity A' and 'Entity B'), but the document addresses only one of them, failing to provide the necessary context for both.",
  
  "The user query requires a relationship or comparison between entities, but the document discusses them in isolation without establishing the link or contrast.",
  
  "The document is topically irrelevant to the user query."
  
]

[Explanation] {explanation}
\end{promptbox}

% 确保导言区已添加 \usepackage{enumitem}
\begin{promptbox}{Prompt: Extraction}
[System Task] You are an expert Logic Analyst for an SEO Pipeline.

Your objective is to consolidate a list of raw rejection reasons into a concise set of\textbf{Canonical Reasons} to inform the development of future optimization tools.

[Merge vs. Segregate Logic]

Determine whether to merge reasons by asking: \textbf{"Would the exact same coding logic or prompt operation fix both issues?"}

1.\textbf{MERGE: Operational Equivalence}

   - *Example:* "Missing price" AND "Missing date".
   
   - *Logic:* Both imply an \textbf{"Insert specific data"} operation. Even though the data differs, the mechanism (Insertion) is the same. Merge them.
   
   - *Example:* "Cookie banner noise" AND "Navigation menu noise".
   
   - *Logic:* Both imply a \textbf{"Delete/Clean pattern"} operation. Merge them.

2. \textbf{SEGREGATE: Operational Distinctness}

   - *Example:* "Content is too short" vs. "Content is off-topic".
   
   - *Logic:* - "Too short" requires \textbf{Expansion} (Generating new context).
   
     - "Off-topic" requires \textbf{Replacement} (Deleting and writing new topic).
     
     - These are different operations. Keep them separate so we can decide later if we need one tool or two.
     
   - *Example:* "Structure is messy (prose)" vs. "Table is broken (data)".
   
     - *Logic:* Prose needs \textbf{Segmentation} . Tables need \textbf{Schema Reconstruction} . Different mechanisms. Keep separate.

[Instruction on Output Format] 

Return the merged list as a single, valid JSON array of strings. Do not use \"json\" or add explanations.

[Original Rules] {rules}

[Merged Rules JSON]
\end{promptbox}

% 确保导言区已添加 \usepackage{enumitem}
\begin{promptbox}{Prompt: Filter}
[Persona] You are a technical writer specializing in creating context-independent documentation.

[Task] Analyze the following rule. Your goal is to remove any part of the rule that makes it dependent on a specific user "query", "question", or "input". The rewritten rule should state a general principle.

- If the rule contains a general principle AND a reference to a query, remove only the query reference. - If the entire rule is ONLY about how to handle a query (e.g., "The document should directly answer the query."), the principle is not general. In this case, you should return an empty string.

[Polarity] Keep the original polarity. If the rule is written as an avoidance/negative statement (e.g., "do not", "should not", "avoid", "lacks"), do NOT flip it into a positive requirement. Preserve the “avoid/should not/lacks” wording when possible.

[Examples] - Input Rule: "The document should provide specific facts and data relevant to the user's query." - Output JSON: "modified rule": "The document should provide specific facts and data."

- Input Rule: "The source must be recent and directly answer the question." - Output JSON: "modified rule": "The source must be recent."

- Input Rule: "The text must be authoritative." - Output JSON: "modified rule": "The text must be authoritative."

- Input Rule: "Directly answer the user's question." - Output JSON: "modified rule": ""

[Instruction on Output Format] Return a single, valid JSON object with one key: "modified rule". The value should be the modified string.

[Input Rule] "{rule}"

[Output JSON]

\end{promptbox}

\subsection{Query Generation}
\label{ssec:query_generation_apx}
This section details the implementation of the 5 specialized prompt templates within the query generator module. Each template is meticulously engineered to simulate authentic user search behaviors—spanning profile extraction, intent classification, and domain relevance filtering—while strictly enforcing persona-specific vocabulary and semantic consistency during the query generation pipeline.

% 确保导言区已添加 \usepackage{enumitem}
\begin{promptbox}{Prompt: Keyword and Persona Extraction}
    
    % --- Role & Goal ---
    You are a Senior Semantic SEO Specialist and User Intent Analyst. Your goal is to reverse-engineer the search intent behind a specific webpage title to uncover high-value keyword clusters and precise audience segments.
    
    \vspace{0.5em}
    % --- Task ---
    \textbf{TASK:} \\
    Analyze the title to identify its core topics and assess the types of users who are likely to be interested in those topics. Crucially, you must expand these topics to include:
    
    \vspace{0.5em}
    % --- Analysis Requirements ---
    \textbf{ANALYSIS REQUIREMENTS:}
    
    \begin{enumerate}[leftmargin=1.5em, noitemsep, topsep=0pt]
        \item \textbf{Topic/Keyword Extraction:}
        \begin{itemize}[leftmargin=2em, noitemsep, topsep=0pt]
            \item \textbf{Core Keywords:} The primary search terms.
            \item \textbf{LSI/Synonyms:} Synonyms and semantically related terms (Latent Semantic Indexing) that add context.
            \item \textbf{Keyword Phrases:} A specific, multi-word search query that reflects clear user intent.
            \item \textbf{Ambiguity:} If a keyword carries multiple meanings (polysemy), appropriately expand it by adding necessary modifiers to strictly define its specific meaning based on the web content. (e.g., clarify "Jaguar" to "Jaguar Car" or "Jaguar Animal").
        \end{itemize}
        
        \item \textbf{Persona Profiling:}
        \begin{itemize}[leftmargin=2em, noitemsep, topsep=0pt]
            \item Identify distinct user personas
            \item For each persona, identify:
            \begin{itemize}[leftmargin=2.5em, noitemsep, topsep=0pt]
                \item \textbf{Role/Demographic:} Specific groups (e.g., Infants, Children, Teens, Adults, Seniors, Pregnant women).
                \item \textbf{Description:} A concise overview of the persona's circumstances, offering general context without referencing the keyword specifically.
            \end{itemize}
        \end{itemize}
    \end{enumerate}
    
    \vspace{0.5em}
    % --- Output Format ---
    \textbf{OUTPUT FORMAT:} \\
    Please present your analysis in the following JSON format:
    
\begin{verbatim}
{{
    "keyword_cluster": {{
    "core": ["keyword1", "keyword2"],
    "lsi_synonyms": ["synonym1", "synonym2"],
    "keyphrases": ["phrase1", "phrase2"]
    }},
    "target_personas": [
    {{
        "name": "Persona Name",
        "description": "Persona Description"
    }}
    ]
}}
\end{verbatim}
\end{promptbox}

\begin{promptbox}{Prompt: Intent Classification}
    
    % --- Role & Task ---
    You are a Senior SEO Specialist and Google Search Quality Rater. Your task is to analyze a website's \textbf{Title} and \textbf{Content Summary} to determine which User Search Intents this page satisfies.
    
    \vspace{0.5em}
    % --- The 4 Search Intents ---
    \textbf{THE 4 SEARCH INTENTS (DEFINITIONS):} \\
    You must select from the following four categories based on the user's goal:
    
    \begin{enumerate}[leftmargin=1.5em, noitemsep, topsep=0pt]
        \item \textbf{Informational (KNOW):} The user wants to learn something, find an answer, or solve a problem. (e.g., Guides, Tutorials, Definitions).
        \item \textbf{Commercial Investigation (COMPARE):} The user is in the consideration phase, researching products/services but not ready to pay yet. (e.g., Reviews, Comparisons, "Best of" lists).
        \item \textbf{Transactional (DO):} The user is ready to take a specific action, usually involving a conversion. (e.g., Buy, Subscribe, Download, Register).
        \item \textbf{Navigational (GO):} The user is looking for a specific website or page. (e.g., Login pages, Contact pages, About Us).
    \end{enumerate}
    
    \vspace{0.5em}
    % --- Task Instructions ---
    \textbf{TASK INSTRUCTIONS:}
    \begin{enumerate}[leftmargin=1.5em, noitemsep, topsep=0pt]
        \item Analyze the provided \textbf{Title} and \textbf{Summary}.
        \item Evaluate the page against \textbf{ALL 4 intents}.
        \item \textbf{Filter Logic (Crucial):} Only assign an intent if the content provides a \textbf{substantial and satisfying solution} for that intent.
        \begin{itemize}[leftmargin=2em, noitemsep, topsep=0pt]
            \item \textit{Example:} A blog post with a small "buy now" link in the footer is \textbf{NOT} Transactional. It is Informational.
            \item \textit{Example:} A product page with a short description is primarily Transactional, but NOT Informational (since it lacks depth).
        \end{itemize}
        \item \textbf{Output:} Return the applicable intents (can be multiple) with a confidence score and reasoning.
    \end{enumerate}
    
    \vspace{0.5em}
    % --- Output Format ---
    \textbf{OUTPUT FORMAT:} \\
    List the suitable intents and provide a short reasoning based on why they were NOT excluded.
    
    \vspace{0.3em}
    \textbf{Output Format:}
    \begin{itemize}[leftmargin=1.5em, noitemsep, topsep=0pt]
        \item \textbf{Intents:} [List of intents chosen from Informational, Commercial Investigation, Transactional, Navigational]
        \item \textbf{Reasoning:} [Brief reasoning explaining the primary goal of the content and any secondary goals based on specific keywords or content features.]
    \end{itemize}

\end{promptbox}

\begin{promptbox}{Prompt: Query Generation}
    
    % --- Role & Goal ---
    You are a Realistic Search Engine User Simulator. Your goal is to mimic the exact search behavior, vocabulary, and intent of a specific user segment.
    
    \vspace{0.5em}
    % --- Input Context ---
    \textbf{INPUT CONTEXT:}
    \begin{itemize}[leftmargin=1.5em, noitemsep, topsep=0pt]
        \item \textbf{Focus Keyword:} \texttt{\{keyword\}}
        \item \textbf{User Persona Profile:}
        \begin{itemize}[leftmargin=1.5em, noitemsep, topsep=0pt]
            \item \textbf{Role/Demographics:} \texttt{\{persona\_name\}} (e.g., Senior Citizen, Python Developer, Busy Mom)
            \item \textbf{User Description:} \texttt{\{persona\_description\}} (Current situation, experience level, environment)
        \end{itemize}
    \end{itemize}
    
    \vspace{0.5em}
    % --- Task ---
    \textbf{TASK:} \\
    Generate exactly 5 search queries related to \texttt{"\{keyword\}"} for EACH of the following applicable search intents.
    
    \vspace{0.5em}
    % --- Intent Categories ---
    \textbf{APPLICABLE SEARCH INTENT CATEGORIES \& GUIDELINES:} \\
    \texttt{\{intent\_section\}}
    
    \vspace{0.5em}
    % --- Output Format ---
    \textbf{OUTPUT FORMAT:} \\
    Return a JSON object with arrays for each intent type. \\
    For intent types NOT listed above, return an empty array \texttt{[]}.
    
% 这里使用 verbatim 环境来保持缩进
\begin{verbatim}
{{
    "navigational": ["query1", ...] or [],
    "informational": ["query1", ...] or [],
    "commercial": ["query1", ...] or [],
    "transactional": ["query1", ...] or []
}}
\end{verbatim}
\end{promptbox}

\begin{promptbox}{Prompt: Deduplication}
    You are an expert Semantic Query Optimizer and Data Cleaner. Your task is to analyze a list of user search queries and remove duplicates based on \textbf{User Intent} and \textbf{Semantic Equivalence}.
    
    \vspace{0.5em}
    \textbf{\# Core Objective}
    
    Reduce redundancy in the provided list while preserving every distinct user intent, specific entity (products/dates), and action type.
    
    \vspace{0.5em}
    \textbf{\# Deduplication Logic (Step-by-Step)}
    
    \begin{enumerate}[nosep, leftmargin=*]
        \item \textbf{Exact \& Morphological Duplicates:}
        \begin{itemize}[nosep, leftmargin=1.5em]
            \item Remove queries that are identical strings (case-insensitive).
            \item Remove queries that are merely reordered variations of keywords without changing the meaning (e.g., "interior 2021 g-wagon" vs. "2021 g-wagon interior").
        \end{itemize}
        
        \item \textbf{Semantic Equivalence (Synonyms):}
        \begin{itemize}[nosep, leftmargin=1.5em]
            \item Merge queries where keywords are swappable synonyms in the specific context.
            \item \textit{Example:} "Luxury cabin" $\approx$ "Premium interior".
            \item \textit{Example:} "Specs" $\approx$ "Specifications" $\approx$ "Features".
            \item \textit{Example:} "Buy" $\approx$ "Purchase".
        \end{itemize}
        
        \item \textbf{Representative Selection (The "Canonical" Query):}
        \begin{itemize}[nosep, leftmargin=1.5em]
            \item When a group of similar queries is identified, select \textbf{one} representative query to keep.
            \item \textbf{Priority for selection:}
            \begin{enumerate}[nosep, leftmargin=1.5em]
                \item The most grammatically natural and fluent option.
                \item The most specific/descriptive option (provided it covers the intent of the others).
                \item The option that uses standard industry terminology.
            \end{enumerate}
        \end{itemize}
    \end{enumerate}
    
    \vspace{0.5em}
    \textbf{\# Strict Exclusion Rules (Do NOT Merge)}
    
    \begin{enumerate}[nosep, leftmargin=*]
        \item \textbf{Distinct Entities:} Do not merge queries referring to different models, versions, or years (e.g., "2021 model" is NOT the same as "2023 model"; "G-Class" is NOT the same as "GLE").
        \item \textbf{Distinct Intents:} Do not merge different stages of the user journey.
        \begin{itemize}[nosep, leftmargin=1.5em]
            \item \textit{Transactional:} "Buy..."
            \item \textit{Informational:} "Reviews of...", "Specs of..."
            \item \textit{Navigational:} "Official site..."
            \item \textit{Comparison:} "Model A vs Model B"
        \end{itemize}
    \end{enumerate}
    
    \vspace{0.5em}
    \textbf{\# Output Format}
    
    Return only the final deduplicated list of strings in valid JSON format: \texttt{["query1", "query2", ...]}
\end{promptbox}

\begin{promptbox}{Prompt: Domain Filtering}
    
    % --- Role & Goal ---
    You are an SEO Relevance Expert and Topic Cluster Specialist. Your goal is to filter a list of search queries based on their potential alignment with a specific document.
    
    \vspace{0.5em}
    % --- Core Objective ---
    \textbf{THE CORE OBJECTIVE:} \\
    You must act as a \textbf{Broad Filter}. Your job is NOT to find only the "perfect" matches. Your job is to \textbf{remove noise}.
    \begin{itemize}[leftmargin=1.5em, noitemsep, topsep=0pt]
        \item \textbf{Default Action:} KEEP the query.
        \item \textbf{Exception:} REMOVE only if the query is definitely irrelevant.
    \end{itemize}
    
    \vspace{0.5em}
    % --- Retention Rules ---
    \textbf{1. RETENTION RULES (When to KEEP):}
    \begin{itemize}[leftmargin=1.5em, noitemsep, topsep=0pt]
        \item \textbf{Same Domain:} Keep all queries related to the document's general industry or topic (e.g., if Doc is about "SEO", keep "digital marketing" queries).
        \item \textbf{Low Relevance / Tangential:} Keep queries that are loosely related. If we could add a paragraph to the document to answer this query, KEEP IT.
        \item \textbf{Broad/Vague:} Keep broad queries (e.g., "what is marketing") if the document covers a specific aspect of that topic.
        \item \textbf{Competitor Comparisons:} Keep queries comparing the subject to alternatives.
    \end{itemize}
    
    \vspace{0.5em}
    % --- Exclusion Rules ---
    \textbf{2. EXCLUSION RULES (When to REMOVE):} \\
    Remove a query \textbf{strictly} if it meets one of these conditions:
    
    \vspace{0.3em}
    \textbf{• Semantic Mismatch (Polysemy):} The query uses a word that appears in the document but refers to a completely different concept/industry.
    \begin{itemize}[leftmargin=2em, noitemsep, topsep=0pt]
        \item \textit{Example (Syndication):} Document is about "Media Content Syndication".
        \item \textbf{KEEP:} "content distribution platforms" (Same domain).
        \item \textbf{REMOVE:} "bank loan syndication process" (Finance domain - completely different audience).
        \item \textit{Example (Python):} Document is about "Python Programming".
        \item \textbf{REMOVE:} "types of python snakes in florida" (Zoology domain).
    \end{itemize}
    
    \vspace{0.3em}
    \textbf{• Wrong Industry/Intent:} The user searching for this has \textbf{zero} overlap with the target audience of the document.
    \begin{itemize}[leftmargin=2em, noitemsep, topsep=0pt]
        \item \textit{Example:} Doc is about "Enterprise CRM Software"; Query is about "Free Minecraft skins" → REMOVE.
    \end{itemize}
    
    \vspace{0.5em}
    % --- Decision Heuristic ---
    \textbf{3. DECISION HEURISTIC:} \\
    Ask yourself: \textbf{"Could the person searching for [Query] potentially find value in [Document Content]?"} \\
    \begin{itemize}[leftmargin=1.5em, noitemsep, topsep=0pt]
        \item If YES (even "1\%" chance): \textbf{KEEP}.
        \item If NO (impossible): \textbf{REMOVE}.
    \end{itemize}
    
    \vspace{0.5em}
    % --- Output Format ---
    \textbf{OUTPUT FORMAT:} \\
    The output queries must be equal to the input! Return the result in JSON format with two distinct lists:

\begin{verbatim}
{
  "relevant_queries": ["list", "of", "queries", "to", "keep"],
  "filtered_queries": ["list", "of", "removed", "queries"]
}
\end{verbatim}

\end{promptbox}

\subsection{Search Failure Analysis}
This section presents the prompt design for search failure analysis, a critical diagnostic step in the optimization pipeline. 
\begin{promptbox}{Prompt: Search Failure Analysis} 
    
    % --- Role & Goal ---
    You are a Search Failure Analyst with expertise in document quality assessment.
    Your task is to determine the \textbf{ROOT CAUSE} of why the Target document was \textbf{NOT} cited while the Competitor was.
    
    \vspace{0.5em}
    \textbf{Query:} \texttt{"\{query\}"}
    
    \vspace{0.5em}
    % --- Inputs ---
    === COMPETITOR DOCUMENT (Winner - Was Cited) === \\
    \texttt{\{competitor\}} \\
    === END COMPETITOR ===
    
    \vspace{0.5em}
    === TARGET DOCUMENT (Loser - Not Cited) === \\
    \texttt{\{target\}} \\
    === END TARGET ===

    \vspace{0.5em}
    \texttt{\{failure\_taxonomy\}}
    
    \vspace{0.5em}
    % --- Instructions ---
    \textbf{ANALYSIS INSTRUCTIONS:}
    \begin{enumerate}[leftmargin=1.5em, topsep=0pt, partopsep=0pt, itemsep=0pt]
        \item \textbf{Compare} the two documents carefully against the query
        \item \textbf{Identify} the PRIMARY reason the Target failed
        \item \textbf{root\_cause MUST be EXACTLY one of}: \\
        {\small\ttfamily % 使用小一号字体并允许换行
        PARSING\_FAILURE, CONTENT\_TRUNCATED, DATA\_INTEGRITY, WEB\_NOISE, \\
        LOW\_SIGNAL\_RATIO, LOW\_INFO\_DENSITY, MISSING\_INFO, STRUCTURAL\_WEAKNESS, \\
        SEMANTIC\_IRRELEVANCE, ATTRIBUTE\_MISMATCH, BURIED\_ANSWER, \\
        NON\_FACTUAL\_CONTENT, TRUST\_CREDIBILITY, OUTDATED\_CONTENT, UNKNOWN}
        \item \textbf{Be Specific} in explanation - identify WHAT specifically is missing or wrong
    \end{enumerate}
    
    \vspace{0.5em}
    % --- Severity Guide ---
    \textbf{SEVERITY GUIDE:}
    \begin{itemize}[leftmargin=1.5em, noitemsep, topsep=0pt]
        \item \texttt{critical}: Document is fundamentally unusable (parsing failure, completely irrelevant)
        \item \texttt{high}: Major issue requiring significant changes (truncation, buried answers, missing key info)
        \item \texttt{medium}: Moderate issue requiring targeted fixes (structure, density)
        \item \texttt{low}: Minor issue requiring polish (formatting, minor noise)
    \end{itemize}

    \vspace{0.5em}
    \textbf{CRITICAL:} Output \texttt{root\_cause} as \textbf{EXACT} category name (e.g., \texttt{"MISSING\_INFO"}, not \texttt{"Missing Information"})

    \vspace{0.5em}  
    \texttt{\{format\_instructions\}}

\end{promptbox} \label{fig:failure_analysis_prompt}

\subsection{Tools Selection Prompt}
This section provides prompt template for the Strategic Tool Selection module, designed to map diagnosed failure modes to the most effective optimization utilities
\begin{promptbox}{Prompt: Tools Selection}
    You are the GEO Optimization Controller. Select the best tool based on the Diagnosis and Policy.
    
    \textbf{ Inputs}
    \begin{itemize}[itemsep=0pt, parsep=0pt, topsep=0pt, partopsep=0pt]
        \item \textbf{Query}: \{query\}
        \item \textbf{Root Cause Diagnosis}: \{diagnosis\_cause\} (\{diagnosis\_explanation\})
        \item \textbf{History}: \{history\_context\}
    \end{itemize}
    
    \textbf{ Policy}
    \{policy\_guidelines\}
    
    \textbf{ Target Document (Indexed)}
    \{target\_content\}
    
    \textbf{ Available Tools}
    \{tool\_descriptions\}
    
    \textbf{ Task}
    Select the tool to resolve the `\{diagnosis\_cause\}'.
    Identify the specific [CHUNK\_ID] that needs modification.
    
    \textbf{ Constraints}
    \begin{itemize}[itemsep=0pt, parsep=0pt, topsep=0pt, partopsep=0pt]
        \item \textbf{Targeting:} You must include \texttt{"target\_chunk\_index": <int>} in your output.
        \item \textbf{Injection Rule:} Include [\texttt{"target\_content"}, \texttt{"context\_before"}, \texttt{"context\_after"}, \texttt{"core\_idea"}, \texttt{"previous\_modifications"}] in \texttt{tool\_arguments} with empty string values (\texttt{""}). The system will populate these.
    \end{itemize}
    
    \{format\_instructions\}
\end{promptbox}

\subsection{Tools Prompt} \label{fig:tool_selection_prompt}

This section provides the implementation details for the \textbf{9 specialized optimization tools}. Each template is engineered to address specific taxonomic failures—spanning information density, structural clarity, and persuasiveness—while strictly enforcing context preservation and semantic integrity during document transformation.

\begin{promptbox}{Example: Intent Realignment Transformation}
    \textbf{Query:} ``How to reduce carbon footprint?'' \\
    
    \textbf{Original:} In today's rapidly changing world, environmental concerns are paramount... \\
    \textbf{Optimized:} \textbf{Reducing your carbon footprint} can be achieved through three primary strategies: transportation changes, energy efficiency, and consumption habits...
    
\end{promptbox}

    \begin{promptbox}{Module: History Section }  
        
        \textbf{PREVIOUS MODIFICATIONS (MUST PRESERVE):} \\ 
        \texttt{\{previous\_modifications\}}
    
    \end{promptbox} 
    
    \begin{promptbox}{Prompt: BLUF Optimization Tool} 
    
        % --- Role & Goal ---
        You are a UX Writer for Information Retrieval.     
        Surface the key takeaway immediately using the BLUF (Bottom Line Up Front) principle.

        % --- Context Variables --- 
        \textbf{Key Takeaway to Surface:} \texttt{\{key\_takeaway\}} \\
        \texttt{\{context\_section\}} \\
        \textbf{Core Idea:} \texttt{\{core\_idea\}} \\
        \texttt{\{history\_section\}}

        % --- Instructions --- 
        \textbf{SPECIFIC INSTRUCTIONS:} 
        \begin{enumerate}[leftmargin=1.5em, topsep=0pt, itemsep=0pt]
            \item Generate the HTML content for a summary box. 
            \item Summarize the ``\texttt{\{key\_takeaway\}}'' in one direct, factual sentence using \texttt{<strong>}.  
            \item Do \textbf{NOT} output the existing content, \textbf{ONLY} the new summary content. 
        \end{enumerate}
    
        \vspace{0.5em}
    
        % --- Output Format Constraint ---
        \textbf{OUTPUT FORMAT:}

            \texttt{[HTML Content for the BLUF Summary Div ONLY]} \\ 
            \\
            \texttt{---MODIFICATION\_SUMMARY---} \\
            \texttt{- [Change 1]}

    \end{promptbox}  
    \label{fig:bluf_prompt}

    \begin{promptbox}{Prompt: Content Relocation Tool}  
    
        % --- Role & Goal ---
        You are a Content Restructuring Expert.  
        The document has \textbf{CRITICAL} information buried deep (truncated from view). 
        Your job is to \textbf{SURFACE} this information by creating a prominent summary section at the TOP. 
        
        \vspace{0.5em}
        
        % --- Context ---
        \textbf{Context} 
        \begin{itemize}[leftmargin=1.5em, topsep=0pt, itemsep=0pt]
            \item \textbf{User Query:} \texttt{\{query\}}
            \item \textbf{Core Idea:} \texttt{\{core\_idea\}} 
            \item \textbf{Problem:} Important information is truncated and invisible. 
            
        \end{itemize} 

        \textbf{Hidden Content Summary(from truncated section)}  
        
        \texttt{\{hidden\_summary\}}   

        \texttt{\{context\_section\}} 
        
        \texttt{\{history\_section\}}

        \texttt{\{preservation\_rules\}} 
        
        \vspace{0.5em}   

        % --- Step 1: Style Analysis --- 
        \textbf{STEP 1: Analyze Content Style}
        
        First, analyze the target content to determine its style:

        \begin{itemize}[leftmargin=1.5em,  topsep=0pt,partopsep=0pt, itemsep=0pt]   
            \item \textbf{Formal/Academic:}  Research papers, official docs, business reports $\to$ Use ``Summary'' or ``Key Findings''  
            \item \textbf{Technical/Documentation:} API docs, tutorials, technical guides  $\to$ Use ``Quick Answer'' or ``Key Points''  
            \item \textbf{Blog/Casual:} Personal blogs, forums, community posts $\to$ Use ``TL;DR'' or ``The Short Version''   
            \item \textbf{News/Editorial:} News articles, reviews, editorials $\to$ Use ``Highlights'' or ``At a Glance''  
            \item \textbf{E-commerce:} Product pages, listings $\to$ Use ``Quick Facts'' or ``Overview'' 
        \end{itemize}

        \vspace{0.5em}

        % --- Step 2: Create Summary ---
        \textbf{STEP 2: Create Summary Section} 
        \begin{enumerate}[leftmargin=1.5em, topsep=0pt,partopsep=0pt, itemsep=0pt]  
            \item \textbf{Choose the most appropriate header} based on the content style detected above.   
            \item \textbf{Synthesize} the hidden content summary into 2-4 bullet points that DIRECTLY answer the query.  
            \item \textbf{Use semantic HTML:} Wrap in \texttt{<div class="geo-summary-box">}, use \texttt{<h3>} for header, \texttt{<ul><li>} for points. 
            \item \textbf{Match the tone:} The summary should feel native to the page - formal content gets formal language, casual content gets casual language.
            \item \textbf{Preserve ALL original content} below this new section - DO NOT delete anything. 
        \end{enumerate}

        \textbf{STEP 2: Style Matching Rules }  
        \begin{enumerate}[leftmargin=1.5em, topsep=0pt, itemsep=0pt]
            \item Be concise but include specific facts, numbers, entities from the hidden summary  
            \item Mirror the vocabulary and tone of the original content 
            \item The new section should blend seamlessly with the existing page  
        \end{enumerate}

        % --- Target Input ---
        === TARGET CONTENT (Current Visible) === \\
        \texttt{\{target\_content\}} \\
        === END TARGET CONTENT ===

        \vspace{0.5em}

        % --- Output Format ---
        \textbf{OUTPUT FORMAT:}

            \texttt{[New Summary Section with style-appropriate header]} \\
            \texttt{[Original Content Preserved Below]} \\
            \\
            \texttt{---MODIFICATION\_SUMMARY---} \\
            \texttt{- Detected content style: [style type]} \\
            \texttt{- Added "[chosen header]" section at top with [specific facts surfaced]}

    \end{promptbox} 
    \label{fig:content_relocation_prompt}

    \begin{promptbox}{Prompt: Data Serialization Tool}  
    
        % --- Role & Goal ---
        You are a Data Structuring Agent.  
        Your task is to convert narrative descriptions of data into structured HTML Tables.

        % --- Context Variables ---
        \texttt{\{context\_section\}} \\
        \textbf{Core Idea:} \texttt{\{core\_idea\}} \\
        \texttt{\{history\_section\}} \\
        \texttt{\{preservation\_rules\}}

        % --- Instructions ---
        \textbf{SPECIFIC INSTRUCTIONS:}
        \begin{enumerate}[leftmargin=1.5em, topsep=0pt, itemsep=0pt]
            \item \textbf{Identify Patterns:} Identify repeating data patterns in the text (e.g., ``Product A costs \$10. Product B costs \$20.'').
            \item \textbf{Create Table:} Create an HTML \texttt{<table>} with appropriate \texttt{<thead>} headers to represent this data.
            \item \textbf{CRITICAL:} Do \textbf{NOT} delete the original narrative text completely if it contains nuance/sentiment that fits poorly in a table. Instead, keep the text as a summary \textit{after} the table, or integrate the nuance into a ``Notes'' column. 
            \item \textbf{Completeness:} Ensure every number and entity from the text exists in the table.
        \end{enumerate}

        % --- Target Input ---
        === TARGET CONTENT (HTML Fragment) === \\
        \texttt{\{target\_content\}} \\
        === END TARGET CONTENT ===

        \vspace{0.5em}

        % --- Output Format Constraint ---
        \textbf{OUTPUT FORMAT:}

            \texttt{[HTML Content with Table...]} \\
            \\
            \texttt{---MODIFICATION\_SUMMARY---} \\
            \texttt{- [Change 1]}

    \end{promptbox} 
    \label{fig:data_structuring_prompt}

    \begin{promptbox}{Prompt: Entity Injection Tool} 
    
        % --- Role & Goal ---
        You are a Content Enricher and HTML Editor.  
        Your task is to perform a ``surgical injection'' of missing information into the provided HTML fragment.

        % --- Context Variables (修复了缺少花括号的问题) --- 
        \textbf{Missing Entities to Inject:} \texttt{\{missing\_entities\}} \\  
        \texttt{\{context\_section\}} \\ 
        \textbf{Core Idea:} \texttt{\{core\_idea\}} \\ % <--- 修复了这里
        \texttt{\{history\_section\}} \\
        \texttt{\{preservation\_rules\}}

        % --- Instructions ---
        \textbf{SPECIFIC INSTRUCTIONS:} 

        \begin{enumerate}[leftmargin=1.5em, topsep=0pt,partopsep=0pt, itemsep=0pt]   
            \item \textbf{Locate the Best Spot:} Analyze the HTML content to find the most semantically relevant place.
            \item \textbf{Rewriting:} Rewrite the relevant sentence/paragraph to naturally include the new fact.  
            \item \textbf{Semantic Highlighting:} Wrap the injected entity in \texttt{<strong>} tags.
            \item \textbf{Preserve Structure:} Do \textbf{NOT} remove existing links (\texttt{<a href...>}) or formatting unless necessary for the rewrite. 
            \item \textbf{Output:} Return the FULL modified HTML chunk.
        \end{enumerate}

        % --- Target Input (清理了不可见字符) ---
        === TARGET CONTENT (HTML Fragment) === \\
        \texttt{\{target\_content\}} \\ 
        === END TARGET CONTENT ===

        \vspace{0.5em}

        % --- Output Format Constraint ---
        \textbf{OUTPUT FORMAT:} 
        First output the enhanced HTML content, then the summary.

            \texttt{[Enhanced HTML content...]} \\
            \\
            \texttt{---MODIFICATION\_SUMMARY---} \\
            \texttt{- [Inserted '\{missing\_entities\}']}

    \end{promptbox}    
    \label{fig:entity_injection_prompt}

    \begin{promptbox}{Prompt: Intent Realignment Tool}   
        
            % --- Role & Goal ---
            You are an expert SEO Content Editor. Your goal is to make the content explicitly answer the User Query immediately, without losing original depth.

            \textbf{User Query:} \texttt{"\{user\_query\}"}

            % --- Context Variables ---  
            \texttt{\{context\_section\}} \\  
            
            \textbf{Core idea:} \texttt{\{core\_idea\}} \\ 
            \texttt{\{history\_section\}} \\
            
            \texttt{\{preservation\_rules\}}

            % --- Instructions ---
            \textbf{SPECIFIC INSTRUCTIONS:}  
        
            \begin{enumerate}[leftmargin=1.5em, topsep=0pt,partopsep=0pt, itemsep=0pt]       
                \item Analyze the intent of the query (Informational/Navigational/Commercial).
                \item Rewrite the specific TARGET CONTENT to address this intent in the first sentence.
                \item Remove "fluff" or vague introductions (e.g., "In today's world..."). 
                \item Ensure keywords related to the query appear naturally.
                \item \textbf{CRITICAL}: If the original content contains specific details (dates, names, specs) not related to the query, MOVE them to the end of the chunk, DO NOT DELETE THEM.  
            \end{enumerate}

            % --- Target Input ---
            === TARGET CONTENT (HTML Fragment) === \\  
            \texttt{\{target\_content\}} \\  
            === END TARGET CONTENT ===
    
            \vspace{0.5em}
    
            % --- Output Format Constraint ---
            \textbf{OUTPUT FORMAT:} 
            First output the enhanced HTML content, then the summary.

                \texttt{[Enhanced HTML content...]} \\
                \\
                \texttt{---MODIFICATION\_SUMMARY---} \\
                \texttt{- [Change 1]}

        \end{promptbox}

    \begin{promptbox}{Prompt: Noise Isolator Tool}   
    
    % --- Role & Goal --- 
        You are an HTML Semantic Cleaner. Your job is to tell search engine parsers what is NOISE and what is CONTENT.

        % --- Context Variables ---  
        \texttt{\{context\_section\}} \\  
        \textbf{Core idea:} \texttt{\{core\_idea\}} \\ 
        \texttt{\{history\_section\}} \\
        \texttt{\{preservation\_rules\}}

        % --- Instructions ---
        \textbf{SPECIFIC INSTRUCTIONS:}  
         
        \begin{enumerate}[leftmargin=1.5em, topsep=0pt,partopsep=0pt, itemsep=0pt]       
            \item Analyze the target content. Identify elements that are likely:   
                \begin{itemize}[label=--, noitemsep, leftmargin=1em, topsep=0pt]
                    \item Navigation links
                    \item Cookie warnings
                    \item Sidebar ads / "Read More" links
                    \item Copyright footers
                \end{itemize}
            \item Wrap these elements in \texttt{<aside>}, \texttt{<nav>}, or \texttt{<footer class="geo-noise">} tags.   
            \item Wrap the actual high-value semantic content in \texttt{<article>} or \texttt{<section>}.
            \item \textbf{DO NOT DELETE} the noise (as it might be functionally needed for the webpage UI), just wrap it semantically so parsers can de-prioritize it. 
        \end{enumerate}

        % --- Target Input ---
        === TARGET CONTENT (HTML Fragment) === \\
        \texttt{\{target\_content\}} \\  
        === END TARGET CONTENT ===
    
        \vspace{0.5em}
    
        % --- Output Format Constraint --- 
        \textbf{OUTPUT FORMAT:}

            \texttt{[Semantically Wrapped HTML...]} \\
            \\
            \texttt{---MODIFICATION\_SUMMARY---} \\
            \texttt{- [Change 1]}

    \end{promptbox}

    \begin{promptbox}{Prompt: Static Renderer Simulator}  
        
            % --- Role & Goal ---
            You are a Server-Side Rendering (SSR) Simulator.
            The input content contains JavaScript code, JSON blobs, or "Loading..." placeholders that hide the actual content.
            Your task is to "render" this into static, readable HTML text based on what the JS code implies. 
            
            % --- Context Variables ---  
            \texttt{\{context\_section\}} \\  
            \textbf{Core idea:} \texttt{\{core\_idea\}} \\ 
            \texttt{\{history\_section\}} \\ 
            \texttt{\{preservation\_rules\}}

            % --- Instructions ---
            \textbf{SPECIFIC INSTRUCTIONS:}  
         
            \begin{enumerate}[leftmargin=1.5em, topsep=0pt,partopsep=0pt, itemsep=0pt]       
                \item Look at the JS code or JSON data in the input. 
                \item Extract the actual human-readable content (text, prices, items) from the code logic. 
                \item Output the content as plain, static HTML tags (\texttt{<p>}, \texttt{<ul>}, \texttt{<div>}).
                \item \textbf{REMOVE} the \texttt{<script>} tags and raw JSON.
                \item If the JS implies fetching external data that is NOT present in the chunk, place a placeholder: \texttt{[Data to be fetched]} but do not hallucinate data.
            \end{enumerate}

            % --- Target Input ---
            === TARGET CONTENT (Raw/JS-heavy Chunk) === \\
            \texttt{\{target\_content\}} \\   
            === END TARGET CONTENT ===

            % --- Output Format Constraint ---
            \textbf{OUTPUT FORMAT:}

                \texttt{[Static HTML Content...]} \\
                
                \texttt{---MODIFICATION\_SUMMARY---} \\
                \texttt{- [Change 1]}

        \end{promptbox}

    \begin{promptbox}{Prompt: Structure Optimization Tool} 
        
            % --- Role & Goal ---
            You are an HTML Semantic Architect. Transform the unstructured text into parser-friendly, structured HTML.

            % --- Context Variables ---  
            \texttt{\{context\_section\}} \\  
            
            \textbf{Core idea:} \texttt{\{core\_idea\}} \\  
            \texttt{\{history\_section\}} \\
            
            \texttt{\{preservation\_rules\}}

            % --- Instructions ---
            \textbf{SPECIFIC INSTRUCTIONS:}  
    
            \begin{enumerate}[leftmargin=1.5em, topsep=0pt,partopsep=0pt, itemsep=0pt]       
                \item \textbf{Semantic Hierarchy}: Insert \texttt{<h3>} or \texttt{<h4>} tags (based on context depth) to break up long text blocks.
                \item \textbf{List Conversion}: Detect sentences that list items (e.g., "We have A, B, and C") and convert them into \texttt{<ul><li>} or \texttt{<ol><li>}.
                \item \textbf{Emphasis}: Wrap key entities (defined in the text) with \texttt{<strong>} or \texttt{<em>}.
                \item \textbf{NO TEXT CHANGES}: You are strictly forbidden from changing the wording, tone, or length of the text. ONLY add HTML tags.
            \end{enumerate}

            % --- Target Input ---
            === TARGET CONTENT (HTML Fragment) === \\
            \texttt{\{target\_content\}} \\   
            === END TARGET CONTENT ===

            % --- Output Format Constraint ---
            \textbf{OUTPUT FORMAT:}

                \texttt{[Structured HTML content...]} \\
                
                \texttt{---MODIFICATION\_SUMMARY---} \\
                \texttt{- [Change 1]}

        \end{promptbox}

\subsection{Citation and Answer Generation}   

To assess optimization efficacy across diverse search engine paradigms, we utilize distinct templates for citation-anchored generation:
\begin{itemize}
    \item \textbf{In-Content Citation Generator:} The LLM receives all retrieved documents concurrently, generating the complete response with integrated citations in a single inference step.
    \item \textbf{Attr-first-then-generate Citation Generator:} A staged pipeline comprising: (1) \textit{Content Selection}—extracting and highlighting core evidentiary sentences from each document; (2) \textit{Attribution clustering}—sequencing selected evidence into thematic clusters aligned with the target response structure; and (3) \textit{Structured Generation}—synthesizing content cluster-by-cluster to construct the final answer while maintaining explicit source attribution throughout.
\end{itemize}
These standardized templates ensure a consistent baseline for evaluating the performance gains achieved through our document optimization pipeline.

    \begin{promptbox}{Prompt: In-Content Citation Generator}      
    % --- System Role ---
        \textbf{System Instruction:}

        Write an accurate and concise answer for the given user question, using \textit{only} the provided summarized web search results. The answer should be correct, high-quality, and written by an expert using an \textbf{unbiased and journalistic tone}. The user’s language of choice (e.g., English, Français, Español, Deutsch, or Japanese) should be used.  
        
        The answer logic must be rigorous and defensible.  Every sentence in the answer should be \textbf{\textit{immediately followed}} by an in-line citation to the search result(s) (e.g., \texttt{[index]}). The cited search result(s) should fully support \textit{all} the information in the sentence.When citing several results, use the \texttt{[1][2][3]} format rather than \texttt{[1, 2, 3]}. You can use multiple search results to respond comprehensively while avoiding irrelevant search results.

        % --- Citation Rules --- 
        \textbf{Human:}  
        Question: \texttt{\{query\}} \\ 
        
        Search Results: \texttt{\{source\}}
        
    \end{promptbox}
    
    \label{fig:citation_prompt}

    \begin{promptbox}{Prompt: Attr-first-then-generate Citation Generator}
    
        % --- 第一部分：Content Selection ---
        \textbf{Content Selection:}  
        
        In this task, you are presented with a question and several text chunks (sentences). The goal is to identify the minimal number of chunks that are relevant to answer the question.  
        
        \begin{itemize}[leftmargin=1.5em, topsep=0pt, itemsep=0pt] 
            \item Return \textbf{ONLY} the indices of the selected chunks as a comma-separated list. 
            \item \textbf{IMPORTANT:} The output must be strictly formatted as: \texttt{[chunk\_index1, chunk\_index2, ...]}. 
            \item \textbf{IMPORTANT:} Chunk indices must follow the format \texttt{'doc\_id-chunk\_id'} (e.g., \texttt{0-0}, \texttt{0-1}, \texttt{1-0}). 
            \item \textbf{IMPORTANT:} Select approximately \textbf{5--15 chunks} that are most relevant to the question.   
        \end{itemize}

        Question: \texttt{\{query\}}

        % 模拟 Python 循环生成的文本
        Chunk [\texttt{0-0}]: \texttt{\{content\_of\_chunk\_0\}} \\
        Chunk [\texttt{0-1}]: \texttt{\{content\_of\_chunk\_1\}} \\
        ... \\
        Chunk [\texttt{doc\_id-sent\_id}]: \texttt{\{content\_of\_chunk\_n\}}
        
        \vspace{0.8em}
        Answer (comma-separated chunk indices only):    

        \vspace{1em} 
        \hrule % (可选) 加一条分割线区分两个任务
        \vspace{1em}

        % --- 第二部分：Clustering ---
        \textbf{Clustering:}  

        In this task, you are presented with a question and several passages, where some parts are ``highlighted'' (namely, there are \texttt{\{<highlight\_start>\}} and \texttt{\{<highlight\_end>\}} tokens before and after each such span). Those spans are supposed to have the information needed to answer the question, and the goal is to fuse all those highlights into a single paragraph that answers the question.   

        As an intermediate step, you need to cluster highlights that can be merged into a sentence (namely, each cluster will be later merged into one sentence). Make sure the clusters are in the same order you would then write the corresponding output paragraph.    

        \begin{itemize}[leftmargin=1.5em, topsep=0pt, itemsep=0pt] 
            \item \textbf{IMPORTANT:} make sure there are at least two clusters. 
            \item \textbf{IMPORTANT:} The output must be of the format \texttt{[\{"cluster":[comma-delimited highlights indices]\}]}. 
        \end{itemize}

        Question: \texttt{\{query\}}

        Content: \\  
        \texttt{... text \{<highlight\_start>\} selected span \{<highlight\_end>\} text ...} 
        
        \vspace{0.8em} 
        Answer 

        \vspace{1em} 
        \hrule % (可选) 加一条分割线区分两个任务
        \vspace{1em}

        % --- 第三部分：Clustering ---  
        \textbf{Generation:}

        In this task, you are presented with a question and several passages, where some parts are ``highlighted'' (namely, there are \texttt{\{<highlight\_start>\}} and \texttt{\{<highlight\_end>\}} tokens before and after each such span). You are also presented with a prefix of a paragraph.   
        
        The goal is to generate a paragraph that answers the question, based on the passages. The given prefix is what you generated so far. Your job is to generate the \textbf{next sentence} in the paragraph, that covers all and only the ``highlighted'' spans.  

        \begin{itemize}[leftmargin=1.5em, topsep=0pt, itemsep=0pt]
            \item Make sure it connects well with the prefix.
            \item Ensure that it covers \textbf{all and only} the ``highlighted'' spans. 
            \item Ensure that it properly builds towards a full and coherent response to the question. 
        \end{itemize}

        Question: \texttt{\{query\}}

        Content: \\  
        \texttt{... text \{<highlight\_start>\} selected span \{<highlight\_end>\} text ...}

        Prefix: \texttt{\{prefix\}} \textit{(Note: Empty if generating the first cluster)}

        \vspace{0.8em}
        Answer:
    
    \end{promptbox}  
    
    \label{fig:chunk_prompt}

%%%%%%%%%%%%%%%%%%%%%%%%%%%%%%%%%%%%%%%%%%%%%%%%%%%%%%%%%%%%%%%%%%%%%%%%%%%%%%%
%%%%%%%%%%%%%%%%%%%%%%%%%%%%%%%%%%%%%%%%%%%%%%%%%%%%%%%%%%%%%%%%%%%%%%%%%%%%%%%
% \newpage
% \input{checklist.tex}

\end{document}